\documentclass[a4paper,11pt]{article}
\pdfoutput=1
\usepackage{jheppub}

\usepackage{float}

\usepackage[utf8]{inputenc}

\usepackage{amssymb,amsmath,amsfonts}
\usepackage{mathtools}
\usepackage{mathrsfs}
\usepackage{bbm}
\usepackage{slashed}
\usepackage{nicefrac}

\usepackage{graphicx}
\usepackage[dvipsnames]{xcolor}
\usepackage{array}

\usepackage{hyperref}
\usepackage{xparse}
\usepackage{xspace}

\usepackage{tikz}
\usetikzlibrary{decorations.pathmorphing}
\usetikzlibrary{automata,positioning}

\usepackage{cancel}
\usepackage[normalem]{ulem}

\usepackage{xifthen}
\usepackage{dsfont}
\usepackage[titletoc]{appendix}
\usepackage{booktabs}
\usepackage{units}
\usepackage{soul}

\newcommand{\gettitle}{}
\hypersetup{linkcolor=black
	colorlinks,
	linkcolor={red!75!black},
	citecolor={blue!75!black},
	urlcolor={blue!75!black},
	pdftitle={\gettitle},
	pdfauthor={Mehtar-Tani},
	pdfkeywords={Perturbative QCD} {Small-x},
	bookmarksopen=true,
	bookmarksopenlevel=2,
	bookmarksnumbered=true
}

\def\bs{\boldsymbol} 
\def\del{\partial}
\def\bdel{\bs\partial}

\newcommand{\eqn}[1]{eq.\,\eqref{#1}}

\newcommand{\nn}{\nonumber\\ }

\def\be{\begin{eqnarray*}}
\def\ee{\end{eqnarray*}}
\def\beq{\begin{eqnarray}}
\def\eeq{\end{eqnarray}}

\newcommand{\bea}{\beq \begin{aligned}}
\newcommand{\eea}{\end{aligned}\eeq}

\newcommand{\0}{{\boldsymbol 0}}

\newcommand{\bx}{{\boldsymbol x}}
\newcommand{\by}{{\boldsymbol y}}

\newcommand{\bz}{{\boldsymbol z}}

\newcommand{\bk}{{\boldsymbol k}}
\newcommand{\bq}{{\boldsymbol q}}

\newcommand{\bn}{{\boldsymbol n}}

\newcommand{\btheta}{{\boldsymbol \theta}}

\newcommand{\cP}{{\cal P}}
\newcommand{\cO}{{\cal O}}

\newcommand{\cG}{{\cal G}}
\newcommand{\cU}{{\cal U}}
\newcommand{\cK}{{\cal K}}
\newcommand{\cW}{{\cal W}}

\newcommand{\tr}{{\rm tr}}
\newcommand{\med}{{\rm med}}
\newcommand{\bkg}{{\rm bkg}}

\newcommand{\rme}{{\rm e}}
\newcommand{\rmd}{{\rm d}}

\newcommand{\qhat}{\hat{q}}

\def\abar{\bar\alpha_s}

\newcommand{\vac}{{\rm vac }}

\begin{document}


\title{Color Coherence and the Soft Structure of QCD Jets in Vacuum and the QGP }



\author[a]{Paul Caucal, }
\emailAdd{caucal@subatech.in2p3.fr}
\affiliation[a]{SUBATECH UMR 6457 (IMT Atlantique, Universit\'e de Nantes, IN2P3/CNRS), 4 rue Alfred Kastler, 44307 Nantes, France}

\author[b]{Yacine Mehtar-Tani }
\emailAdd{mehtartani@bnl.gov}
\affiliation[b]{Physics Department, Brookhaven National Laboratory, Upton, NY 11973, USA}



\abstract{We formulate a theoretical framework for the evolution of QCD jets in vacuum and in the quark--gluon plasma through the resummation of large energy logarithms. Exploiting the strong hierarchy between the hard scale of the jet and the energy scale associated with jet energy loss, we show that jet observables near threshold can be formulated in terms of Wilson-line correlators obeying Banfi--Marchesini--Smye (BMS) evolution. In this description, soft radiation resolves the internal color structure of the jet, leading to a hierarchy of non-linear evolution equations that govern the evolution of color coherence and the emergence of decoherent energy loss.
For jets propagating through a QCD medium, we demonstrate that medium-induced interactions modify the boundary conditions of the evolution while leaving its ultraviolet structure unchanged. This separation of scales provides a unified description of vacuum-like radiation, medium-induced energy loss, and color coherence. In the large-$N_c$ limit, the resulting evolution is closely related to the Balitsky--Kovchegov equation of high-energy QCD, allowing concepts from saturation physics to be applied to jet quenching. In particular, the medium coherence angle plays a role analogous to the saturation scale and acquires the same asymptotic scaling behavior under evolution.
Our framework establishes a perturbative foundation for the study of color coherence effects in jet quenching and provides a unified picture of soft jet evolution in vacuum and in dense QCD matter.
}

\keywords{Perturbative QCD, Jets, Jet quenching, Heavy-Ion Collisions, Gluon saturation, Factorization}

\maketitle

\section{Introduction\label{sec:intro}}

The suppression and modification of energetic jets produced in heavy-ion collisions, commonly referred to as jet quenching, provides one of the most direct probes of the quark--gluon plasma created in ultrarelativistic nuclear collisions. Since the first observations of high-$p_T$ hadron suppression and the disappearance of back-to-back correlations at RHIC~\cite{Adcox:2001jp,Adler:2002xw,Adler:2002tq}, followed by measurements of reconstructed jets at the LHC~\cite{Aad:2012vca,Aad:2014bxa,Adam:2015ewa,Khachatryan:2016jfl,CMS:2021vui,ALICE:2023waz}, jet quenching has evolved into a precision tool for studying the transport properties and microscopic structure of hot QCD matter. Comprehensive reviews of the subject can be found in Refs.~\cite{Mehtar-Tani:2013pia,Blaizot:2015lma,Cao:2020wlm,Cao:2024pxc,Mehtar-Tani:2025rty}.

The theoretical description of jet quenching originated from the observation that energetic partons lose energy while traversing dense QCD matter through multiple scattering and induced gluon radiation~\cite{Gyulassy:1990ye,Wang:1991xy,Wang:1992qdg,Kovner:2003zj,Baier:1996kr,Baier:1996sk,Zakharov:1996fv,Zakharov:1997uu}. This picture was subsequently developed into a quantitative framework through higher-twist approaches~\cite{Guo:2000nz,Wang:2001ifa}, opacity expansions, and multiple-soft-scattering formalisms, culminating in modern descriptions of medium-induced radiation and parton cascades~\cite{Blaizot:2012fh,Blaizot:2013vha,Arnold:2008zu,Arnold:2020uzm,Feal:2019xfl,Andres:2020vxs}. Significant progress has also been achieved in understanding transverse momentum broadening and the role of Glauber interactions in jet propagation through matter~\cite{DEramo:2012uzl,Mueller:2016gko,Vaidya:2020cyi}.

A major conceptual development emerged with the realization that jets do not interact with the medium as collections of independent partons. Instead, the medium resolves the internal color structure of the jet only above a characteristic coherence scale. This led to the modern picture of color coherence and decoherence in jet quenching~\cite{Mehtar-Tani:2010ebp,Mehtar-Tani:2011hma,Mehtar-Tani:2011vlz,Casalderrey-Solana:2011ule,Mehtar-Tani:2011lic,Mehtar-Tani:2012mfa,Casalderrey-Solana:2012evi}. In this framework, jet evolution is governed by the competition between vacuum-like splittings and medium-induced interactions, giving rise to characteristic phenomena such as anti-angular ordering, decoherence, and the suppression of wide-angle radiation. These ideas have played a central role in the development of modern descriptions of jet substructure in heavy-ion collisions~\cite{Mehtar-Tani:2016aco,Caucal:2018dla,Caucal:2019uvr,Mehtar-Tani:2021fud,Takacs:2021bpv,Attems:2022otp,Abreu:2024wka}.

Despite these advances, a fundamental challenge remains unresolved: the construction of a systematic all-order theory of jet quenching. In vacuum, the evolution of jets is understood in terms of factorization theorems and renormalization-group equations that resum logarithmically enhanced corrections to all orders in perturbation theory~\cite{Collins:2011zzd,Becher:2014oda}. The phenomenology of inclusive jet production and jet substructure has reached a high level of precision through the resummation of soft and collinear logarithms~\cite{Dasgupta:2016bnd,Kang:2016mcy,Dai:2017dpc,Lee:2024tzc,Dasgupta:2020fwr}. Moreover, observables sensitive to soft radiation outside jets are known to be governed by non-linear evolution equations, notably the Banfi--Marchesini--Smye (BMS) equation~\cite{Banfi:2002hw}, whose relation to Wilson-line evolution and high-energy QCD has been extensively explored~\cite{Hatta:2009nd,Larkoski:2015zka,Becher:2016mmh,Caron-Huot:2015bja,Neill:2016stq}.

By contrast, jet-quenching theory has largely been formulated in terms of transport processes, medium-induced branching, and energy-loss probabilities. While these approaches successfully capture important aspects of jet-medium interactions, they do not naturally expose the underlying renormalization-group structure of the problem. As a consequence, the connection between vacuum resummation, color coherence, and medium-induced dynamics has remained largely implicit.

A significant step toward resolving this issue was recently achieved in Ref.~\cite{Mehtar-Tani:2024smp,Mehtar-Tani:2025xxd}, where inclusive jet production in heavy-ion collisions was reformulated using methods borrowed from effective field theory and open quantum systems. That work showed that medium effects can be encoded in Wilson-line correlators and factorized from the short-distance production process, thereby providing the first systematic framework in which jet quenching can be formulated as a problem of QCD evolution. The present work builds upon this observation and develops the corresponding evolution equations. A similar approach was pursued concurrently at leading order in opacity in Ref.~\cite{Vaidya:2026yfa}, also using the effective field theory developed in Ref.~\cite{Mehtar-Tani:2025xxd}, for a dilute medium whose length is much shorter than the mean free path. By contrast, our work focuses on the renormalization-group structure of soft-collinear resummation in a dense, extended QCD medium.

Our first key insight, motivated by the central role of jet energy loss in jet quenching, is that the threshold limit of the factorization theorem for the inclusive jet cross-section provides the natural framework for the resummation of large soft-collinear logarithms. In this limit, powers of the inverse spectral index are neglected, i.e.\ $N\gg 1$, where $N$ denotes the power governing the steeply falling jet spectrum. The jet function then  refactorizes into the Wilson-line correlators mentioned above, whose renormalization group evolution with the factorization scale is governed by the BMS equation in the large-$N_c$ limit. Evaluating these correlators on a medium state rather than on the vacuum leaves the renormalization group equation unchanged, but modifies its boundary condition because of the hierarchy between the jet energy $E$, the characteristic energy lost through vacuum radiation, $\sim E/N$, and the medium-induced energy-loss scale. The effect of the medium is therefore entirely encoded in the initial condition for the BMS evolution, which can be computed within a separate effective field theory describing dynamics below the onset of the vacuum shower. For a dense, weakly coupled, and extended quark--gluon plasma described by its transverse momentum broadening coefficient $\qhat$ and size $L$, this medium-modified initial condition reproduces the standard Baier-Dokshitzer-Mueller-Peigné-Schiff-Zakharov (BDMPS-Z) gluon spectrum~\cite{Baier:1996kr,Baier:1996sk,Zakharov:1996fv,Zakharov:1997uu}, including the Landau-Pomeranchuk-Migdal (LPM) effect~\cite{Landau:1953um,Migdal:1956tc}, as well as color-coherence effects in the medium-induced radiation pattern generated by a QCD antenna. 

We then discuss the physical implications of this framework, focusing in particular on how BMS evolution modifies this medium-induced initial condition. Our conclusions rely crucially on the close analogy with the high-energy (small-$x$) Balitsky--Kovchegov (BK) and Jalilian-Marian--Iancu--McLerran--Weigert--Leonidov--Kovner (JIMWLK) evolution ~\cite{Balitsky:1995ub,Kovchegov:1999yj,Jalilian-Marian:1997qno,Jalilian-Marian:1997jhx,Kovner:2000pt,Iancu:2000hn,Iancu:2001ad,Ferreiro:2001qy} evolution of the initial condition describing the wave function of a large nucleus (in the context of cold nuclear matter effects). In particular, the soft-collinear operator built from two collinear Wilson lines, which physically represents the energy-loss probability distribution of a color-singlet dipole, is the direct analogue of the dipole $S$-matrix in high-energy scattering. Within this correspondence, the medium (de)coherence angle $\theta_c\sim (\qhat L^3)^{-1/2}$ plays the role of the saturation scale in the initial condition for the nucleus wave-function --- for instance, either in the McLerran-Venugopalan (MV)~\cite{McLerran:1993ni,McLerran:1993ka} or Golec-Biernat-W\"{u}sthoff (GBW)~\cite{Golec-Biernat:1998zce} models ---, marking the transition between the color-transparency and strong-scattering regimes. In the context of jet quenching, this transition also corresponds to the onset of coherent versus incoherent energy loss. The BMS evolution then leads to a scale dependence of $\theta_c$, in close analogy with the $x$ dependence of the saturation scale predicted by the BK equation. As the evolution proceeds, corresponding to increasing jet energy, the critical angle decreases with a scaling law which is identical to that of the increase of the saturation scale with decreasing $x$. Our work therefore reveals a deep connection between two QCD phenomena that have traditionally been viewed as distinct because they arise in very different physical settings: gluon saturation, an initial-state phenomenon in high-energy scattering off a large nucleus, and color coherence, a final-state phenomenon governing jet propagation through a dense medium. We show that these phenomena are in fact two manifestations of the same underlying dynamics. Their unification rests on the central role of QCD color correlations and on the way these correlations are renormalized by quantum fluctuations, which governs both gluon saturation at small $x$ and the onset of color decoherence in jet quenching. The jet-quenching vs.~saturation correspondence and its implications for the phenomenology of high-$p_T$ jet suppression in heavy-ion collisions at RHIC and the LHC are summarized in the companion Letter~\cite{Caucal:2026}.

Our paper is organized as follows. In section~\ref{sec:scale}, we derive the factorization formula for the inclusive jet production cross section in the threshold limit, highlighting the need to refactorize the jet function in terms of Wilson-line correlators along the directions of the collinear partons. These correlators describe the soft-collinear modes of the effective theory, which provide the dominant contribution to jet energy loss through vacuum radiation in this limit. Section~\ref{sec:c-coft-function} is devoted to the one-loop calculation of the soft-collinear operator for a single collinear parton, together with the associated collinear function. Building on these results, section~\ref{sec:BMS-evol} derives the renormalization group equation for the soft-collinear operator and discusses its large-$N_c$ limit, in which the infinite hierarchy closes and reduces to the BMS equation. In section~\ref{sec:med-boundary}, we show that medium effects factorize from the vacuum-like evolution and are entirely encoded in the initial condition of the evolution equations. We also briefly summarize the corresponding initial condition in the BDMPS-Z framework. The interplay between this medium-modified initial condition and the subsequent BMS evolution is investigated in section~\ref{sec:thetac_evol}, which contains the main physical results of this paper. Finally, section~\ref{sec:Wilson-correlators-eft} presents the effective field theory formulation of the medium-modified initial condition, thereby providing a unified framework for describing jet evolution both in the vacuum and in the presence of a dense QCD medium. We conclude by summarizing our main results and discussing several directions for future work.

\section{Factorization of inclusive jet production in the threshold limit \label{sec:scale}}
The inclusive jet production cross section --- both in vacuum (as in proton-proton collisions) and in heavy-ion collisions --- factorizes into a hard function, $H(p_T,\mu)$, and a jet function, $J(\mu, p_T R)$. The hard function encodes the short-distance (early-time) physics associated with the hard scale $\sim p_T$, where $p_T$ is the jet transverse momentum. The jet function, on the other hand, describes the collinear dynamics along the jet axis, governed by the lower scale $\sim p_T R$.
Assuming a small jet radius, $R \ll 1$, this separation of scales justifies the following factorization formula:
\begin{align}\label{eq:fact-pp}
    \frac{\rmd \sigma_{\rm jet}}{\rmd p_T} = \int_0^1 \frac{\rmd z}{z} \ H(p_T/z,\mu) \, J(z,\mu,p_TR)+\cO(R^2)\,.
\end{align}
This formula forms the basis of the small-$R$ resummation of collinear logarithms, which has been instrumental in achieving a quantitative understanding of jet cross sections~\cite{Dasgupta:2014yra,Dasgupta:2016bnd,Kang:2016mcy,Dai:2016hzf}. We have intentionally dropped the flavor dependence to alleviate the notations. $J\equiv J_{{\rm jet}/i}$ and $H\equiv  H_i $ are implied where, $i=$ quark or gluon.

The central assumption underlying this factorization in heavy-ion collisions is that the hard function remains insensitive to long-distance, non-local dynamics. In particular, the short-distance production of a high-$p_T$ parton occurs on time and length scales much smaller than those associated with the medium, and therefore factorizes from soft physics, up to initial-state effects encoded in the parton distribution functions (PDFs), which can differ from those in proton-proton collisions.
A more subtle potential source of factorization breaking arises from soft rescattering effects~\cite{Collins:1988ig,Collins:2007nk,Mulders:2011zt,Catani:2011st,Forshaw:2012bi}, such as Glauber exchanges, which can couple collinear modes within the jet to the underlying event. These interactions may, in principle, modify the color flow and generate correlations between the hard production process and the medium. For a medium composed of a large number of approximately independent participant nucleons, however, such connected correlations are diluted and are parametrically suppressed by inverse powers of the number of binary nucleon-nucleon collisions ($N_{\rm coll}$). They may therefore be neglected at leading order in the large $N_{\rm coll}$ expansion.
In particular, for large nuclei, the hard scattering that produces the jet can be viewed as a localized nucleon-nucleon interaction embedded within a much larger system. In this limit, correlations between the hard process and the surrounding medium are diluted by the large number of independent degrees of freedom. Consequently, invoking an analogy with the molecular chaos hypothesis, one may assume that, at leading power in the system size, the medium is effectively uncorrelated with the primary hard scattering. The jet thus propagates through a QCD medium that acts as an ensemble characterized by bulk properties (e.g., temperature, density), rather than by detailed correlations with the initial hard event.
This separation of scales and loss of correlations justifies the factorized description at leading power, with medium effects entering through modifications of the jet function rather than the hard function. This approach has also enabled the extraction of medium-modified jet functions directly from experimental data~\cite{Qiu:2019sfj}.

Numerical studies of jet cross sections, notably Ref.~\cite{Dasgupta:2016bnd}, have demonstrated that the resummation of $\ln(1/R)$ terms remains phenomenologically important even for moderately small jet radii. These logarithmically enhanced contributions can be systematically resummed within a DGLAP-like evolution framework \cite{Dasgupta:2016bnd,Kang:2016mcy,Lee:2024tzc,Mehtar-Tani:2024mvl}. 
In the small-$R$ limit, there is a parametrically large phase space for radiation emitted at angles between $R$ and $\mathcal{O}(1)$. Such emissions escape the jet cone and reduce the energy of the reconstructed jet from its initial value $E =  p_T/z$ to the measured transverse momentum $p_T$. In this sense, the jet function can be interpreted as describing energy loss in vacuum.

As will be shown below, the enhanced sensitivity to this resummation can be traced to the emergence of an effective double-logarithmic structure of the form $\ln R \, \ln N$, where $N \gg 1$ characterizes the power index of the steeply falling jet spectrum. We argue that the interplay between angular and spectral logarithms is the primary origin of the sizable impact of resummation effects, even when $R$ is not parametrically small. 

More specifically, the NLO correction, enhanced by $\ln R$, is effectively weighted by the steep energy dependence of the hard matrix element, generating an additional logarithm associated with the ratio of energy scales $E \sim p_T$ and $E \sim p_T/N$. Here, $N$ denotes the spectral index defined through $H(E) \sim E^{-N}$, with typical phenomenological values $N \sim 5$--$6$ at the LHC and $N \sim 8$--$10$ ar RHIC energies. The precise definition of $N$ will be given in section~\ref{sub:fact-formula}.
This observation provides the starting point for the resummation framework developed in this work. A similar idea was explored in \cite{Becher:2007ty} in the context of Drell-Yan production near threshold, where the steep falloff of the PDF as $x\to 1$ enhances the soft logarithms. 

This perspective is particularly useful when extending the formalism to jet evolution in a QCD medium, where energy loss mechanisms play a central role. The key ingredient enabling a unified description is the separation of time scales between early-time vacuum-like evolution and late-time medium-induced radiation. This separation allows one to treat both dynamics within a common framework, which is most naturally formulated in the soft limit of QCD, as will be discussed below.

\subsection{Fixed order estimate of the jet function: emergence of threshold logarithms}

Let us illustrate this by analyzing the one loop contribution to the (quark) jet function. Indeed, because of the steepness of the initial jet spectrum, the soft limit of the splitting function dominates, resulting in large double logarithms: $\alpha_s \ln R \ln N$. This can be easily diagnosed at one loop order 
\beq
\frac{\rmd \sigma^{\rm NLO}}{\rmd p_T }  &=&\int_0^1 \frac{\rmd z}{z} J(z,R) \frac{\rmd \sigma^{\rm LO} }{\rmd p_T} (p_T/z)\,,\\
&\approx &\frac{\rmd \sigma^{\rm LO} }{\rmd p_T}   \int_0^1\rmd z \,z^{N-1} \, J(z,R) \,,\\
&= &\frac{\rmd \sigma^{\rm LO} }{\rmd p_T}  \left[1+ \frac{\alpha_sC_F}{\pi} \ln R \int_0^1 \rmd z \,z^{N-1} \, P_{qq}(z)+\mathcal{O}(\alpha_s^2)\right]\,.
\eeq
To get the last line, we have performed one step in the collinear DGLAP evolution of the jet function.

Now, keeping only the dominant contribution near $z\to 1$, we have 
\begin{align}\label{eq:pqq-small-z}
 \int_0^1 \rmd z \,z^{N-1} \, P_{qq}(z) &\sim  \int_0^1 \rmd z \,\frac{  2z^{N-1}}{(1-z)_+} \,\\
 &=  2\frac{\Gamma'(N-1)}{\Gamma(N-1)}\sim - 2(\ln N+\gamma_E)+\mathcal{O}\left(1/N\right)\,,
\end{align}
where $\gamma_E\approx 0.5772$ is the Euler-Mascheroni constant. Note that the color factor has been absorbed into the coupling constant, rather than being included in the splitting function. In addition to the soft contribution, there is a purely collinear term arising from virtual corrections, given by $\frac{3}{2}\,\delta(1-z)$ for the quark channel of the jet function. Upon integration, this term contributes an additive $\frac{3}{2}$ to eq.\,\eqref{eq:pqq-small-z}. This corresponds to the well-known threshold limit of the Altarelli--Parisi splitting function. A numerical evaluation of the exact $z$-integral in eq.\,\eqref{eq:pqq-small-z} indicates that power-suppressed corrections yield $\sim 0.02$ which amounts to an order 1\% correction compared to the leading threshold contribution for $\ln N \simeq \ln 6 \simeq 1.8$ that yields $-3.24$. Therefore, although $\ln N$ is not parametrically large, it nevertheless provides the dominant contribution in practice.

As a result, we obtain 
\beq\label{eq:nlo-mellin}
\frac{\rmd \sigma^{\rm NLO}}{\rmd p_T }  
&\approx &\frac{\rmd \sigma^{\rm LO} }{\rmd p_T}  \left[1- \frac{2\alpha_sC_F}{\pi} \ln \frac{1}{R} \left(\ln N +\gamma_E-\frac{3}{2}\right) +\mathcal{O}(\alpha_s^2)\right]\,.
\eeq
Here, one identifies the emergence of Sudakov double logarithms, which encode the dominance of virtual soft radiation at large angles \cite{Mehtar-Tani:2017web,Mehtar-Tani:2024mvl}. The relevant phase space corresponds to gluon energies in the range $E > \omega > E_{\rm loss} \equiv p_T/N$, where emissions outside the jet cone contribute to the suppression of the jet energy. For softer emissions, $\omega < E_{\rm loss}$, real and virtual contributions largely cancel, leading to no net effect at leading power.

A more complete calculation, to be presented in section~\ref{sec:c-coft-function}, shows the presence of a $\ln^2 N$ term that is:
\beq 
\alpha_s \ln \left(\frac{1}{R}\right) \ln N \quad \to \quad \alpha_s \ln \left(\frac{N}{R}\right) \ln N\,.\label{eq:lnN-logs}
\eeq
It is instructive to estimate the size of this effect numerically. For $R=0.3$, $N=6$, and $\alpha_s=0.2$, we find
$\alpha_s \ln\!\left(N/R\right)\ln N \sim 1.07 \, $.
This clearly demonstrates the need to resum both collinear and soft logarithms. At RHIC energies, the jet spectrum is even steeper, making resummation an even more important ingredient of the theoretical description.

The logarithmic enhancement for $N\gg 1$ identified in this fixed order estimate and the phenomenologically large value of $\ln N$ indicates that the underlying dynamics is more naturally described by BMS evolution rather than DGLAP, upon identifying $\ln(E_{\rm loss}/E) \to \ln N$. This will be the central focus of section~\ref{sec:BMS-evol}. In the presence of a QCD medium, this observation also suggests that the jet function obeys a generalized BMS evolution equation. Medium effects are incorporated through a non-trivial initial condition, which resums soft, medium-induced radiation at frequencies $\omega \lesssim p_T/N$, in contrast to the vacuum boundary condition, which is unity~\cite{Banfi:2002hw}.

\subsection{Simultaneous collinear and collinear-soft factorization}
\label{sub:fact-formula}

\begin{figure}[t]
\centering
\includegraphics[width=12cm]{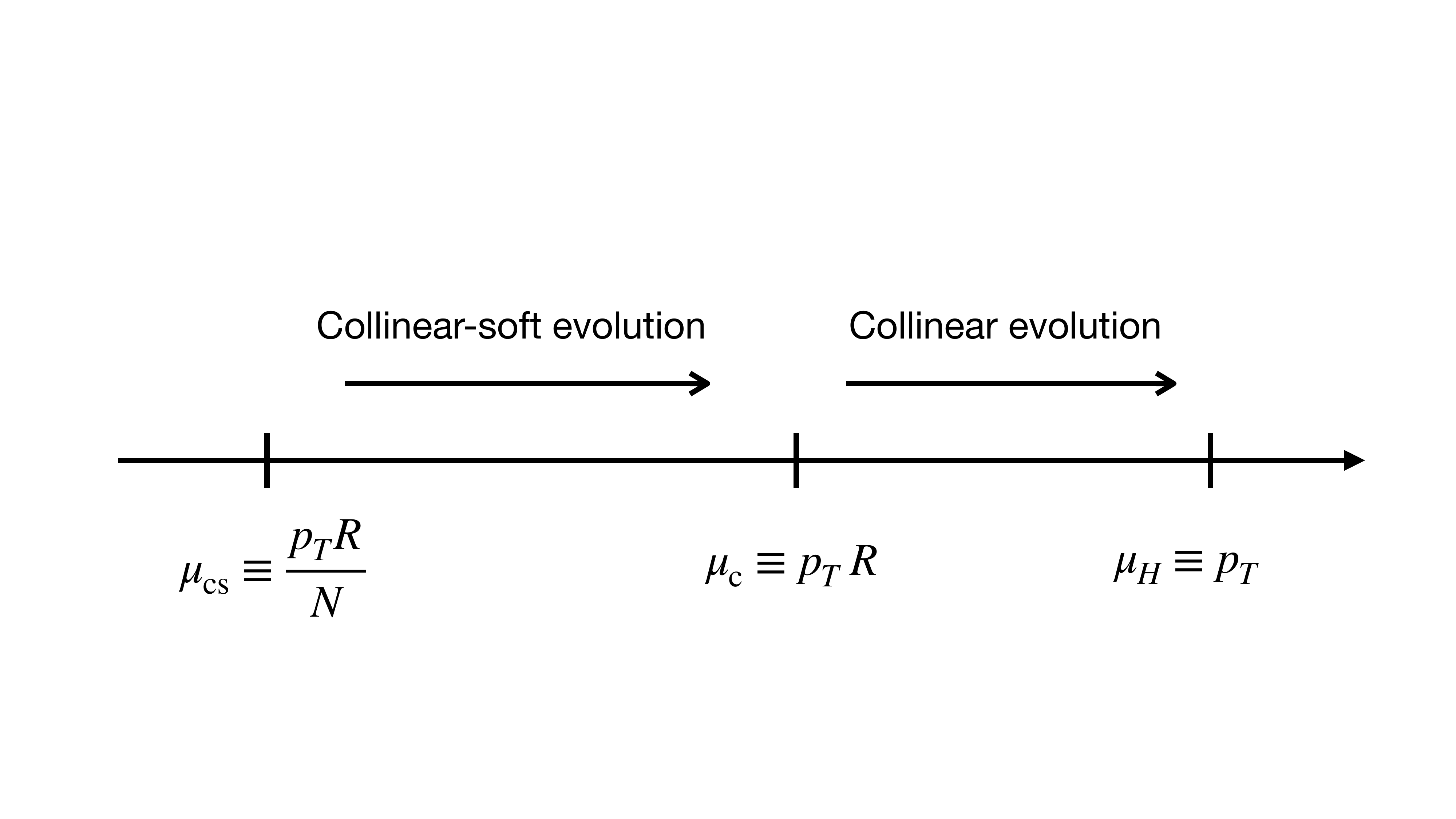}
\caption{The double resummation of $\ln R$ and $\ln N$ is performed via the evolution of the collinear function $C$ from $\mu_{\rm c}= p_T R  $ to $\mu_H=p_T$ and the collinear-soft $S$ from $\mu_{\rm cs}= p_T R/N  $ to $\mu_{\rm c}= p_T R  $. \label{fig:natural-scale-choices}}
\end{figure}

The large $\ln(N)$ logarithms, in addition to the potentially large $\ln(R)$ logarithm call for a factorization theorem which simultaneously account for their joined resummation (see figure~\ref{fig:natural-scale-choices}). This factorization theorem can be obtained by considering the threshold limit of the jet function where powers of $1-z\sim 1/N$ are systematically neglected. The resulting factorization formula then accounts for two kinds of QCD modes.
In addition to collinear modes, which scale (in light-cone coordinates\footnote{In this work we use the following conventions: $k^+=E+k_z$ and $k^-=(E-k_z)/2$} $k \equiv (k^+,k^-,k_\perp)$) as
\beq
k_{\rm coll} \sim p_T\, (1, R^2, R)\,,
\eeq
one must also include collinear-soft modes that describe soft radiation near the jet boundary at angle $R$,
\beq
k_{\rm csoft} \sim p_T\, \beta\, (1, R^2, R)\,,
\eeq
where the expansion parameter is
\beq 
\beta \equiv \frac{E_{\rm loss}}{p_T} \sim \frac{1}{N}\,.
\eeq

Our starting point is the factorization formula in the small-$R$ limit given by eq.\,\eqref{eq:fact-pp} and valid up to power corrections in $R^2$. As mentioned already, the hard factor $H_i$ describes the short-distance production of the final state parton $i$, but note that in hadronic collisions, $H_i$ also includes the convolution with collinear parton distribution functions.\footnote{Compared to the factorization formula presented in~\cite{Mehtar-Tani:2025xxd,Vaidya:2026yfa}, we also absorb the global soft function into the definition of $H$, since it is not modified by the presence of the medium and will therefore not be relevant for our discussion.} A central assumption is that $H_i$ is a steeply falling function of the parton $p_T$. Defining the (local) spectral index as 
\begin{align}
    N&\equiv -\frac{\rmd \ln H(p_T,\mu)}{\rmd \ln(p_T)}\,,
\end{align}
and $\epsilon=(1-z)p_T$, one can express $H(p_T/z,\mu)$ for $N\gg 1$ in the threshold limit as
\begin{align}
H(p_T/z,\mu)&\simeq H(p_T+\epsilon,\mu)=\frac{1}{(p_T+\epsilon)^N} \,,\\
&= H(p_T,\mu) \times\exp\left(-\frac{N\epsilon}{p_T}\right)+\cO(\epsilon^2/p_T^2)\,. 
\end{align}
It is then convenient to work in Laplace space where the energy convolution in eq.\,\eqref{eq:fact-pp} factorizes as
\begin{align}
    \frac{\rmd \sigma_{\rm jet}}{\rmd p_T }&=\left. H(p_T,\mu)\tilde J(\nu,\mu)\right|_{\nu=N/p_T}\,,
\end{align}
with the Laplace transform of $J$ defined by
\begin{align}
    \tilde J(\nu,\mu)&=\int_0^\infty \rmd \varepsilon \ e^{-\nu\varepsilon}J(1-\varepsilon/p_T,\mu)\,.
\end{align}
Note that $N$ may have a weak dependence on $p_T$ and $\mu$ which we omit in our notation.

Before addressing the non-linear dynamics encoded in the BMS equation, we first consider the threshold limit of eq.\,\eqref{eq:fact-pp}, which is sensitive to radiation outside the jet cone. The corresponding anomalous dimension is, to leading power, insensitive to the internal jet substructure. In this sense, it can be interpreted as arising from the renormalization, or QCD evolution, of the total color charge of the jet. Physically, this contribution is associated with ultraviolet (UV) dynamics, reflected in divergences arising from modes with transverse momentum $k_\perp \to  p_T$. 
We first focus on the contribution associated with the total charge. In the limit $N \to \infty$, both the hard function and the jet function admit a systematic expansion and the jet function re-factorizes as
\begin{align}
J(z,\mu)= \int_0^\infty \rmd\omega \, \delta((1-z)E-\omega)\, S_{\rm soft}(\omega,\mu) C_{\rm coll}(\mu)   \,, \label{eq:J-refact-total}
\end{align}
or equivalently, in Laplace space
\beq \label{eq:factorization}
 \tilde J(\nu,\mu)= \tilde S_{\rm soft}(\nu,\mu)\, C_{\rm coll}(\mu)+\cO(1/N)\,.
 \eeq
The soft function $S_{\rm soft}$ encodes soft radiation emitted outside the jet cone by the leading hard parton, while the collinear function $C_{\rm coll}(\mu)$ resums unresolved collinear fluctuations, which are predominantly virtual. In the approximation where collinear radiation inside the jet is not resolved, $C_{\rm coll}(\mu)$ effectively accounts for the virtual phase space of emissions at angles larger than $R$ (the real radiations are canceled by their virtual counter part at angles smaller than $R$). This corresponds to the approximation adopted in Ref.~\cite{Dai:2017dpc}. 

The factorization in eq.\,(\ref{eq:factorization}) naturally separates the two logarithmic evolutions, as illustrated in figure~\ref{fig:natural-scale-choices}. The collinear-soft function evolves from its natural scale $\mu_{cs}=p_T R/N$ to the collinear scale $\mu_c=p_T R$, resumming logarithms of $N$, while the collinear function evolves from $\mu_c$ to the hard scale $\mu_H=p_T$, resumming logarithms of $R$. Together, these two evolution steps achieve the simultaneous resummation of both logarithms.

However, near the lower boundary of the renormalization group evolution, $\mu \sim p_T/N$, collinear-soft modes begin to resolve the internal jet structure, and the description in terms of a single charge becomes insufficient. In this regime, a more general formulation in terms of multi-Wilson-line operators is required, as will be discussed below. Conversely, a complete resummation of logarithms of $N$ requires the inclusion of resolved collinear emissions within the jet. Accounting for such contributions is also essential for capturing the dynamics of color decoherence, particularly in the presence of a medium. 

\begin{figure}[t]
\centering
\includegraphics[width=12cm]{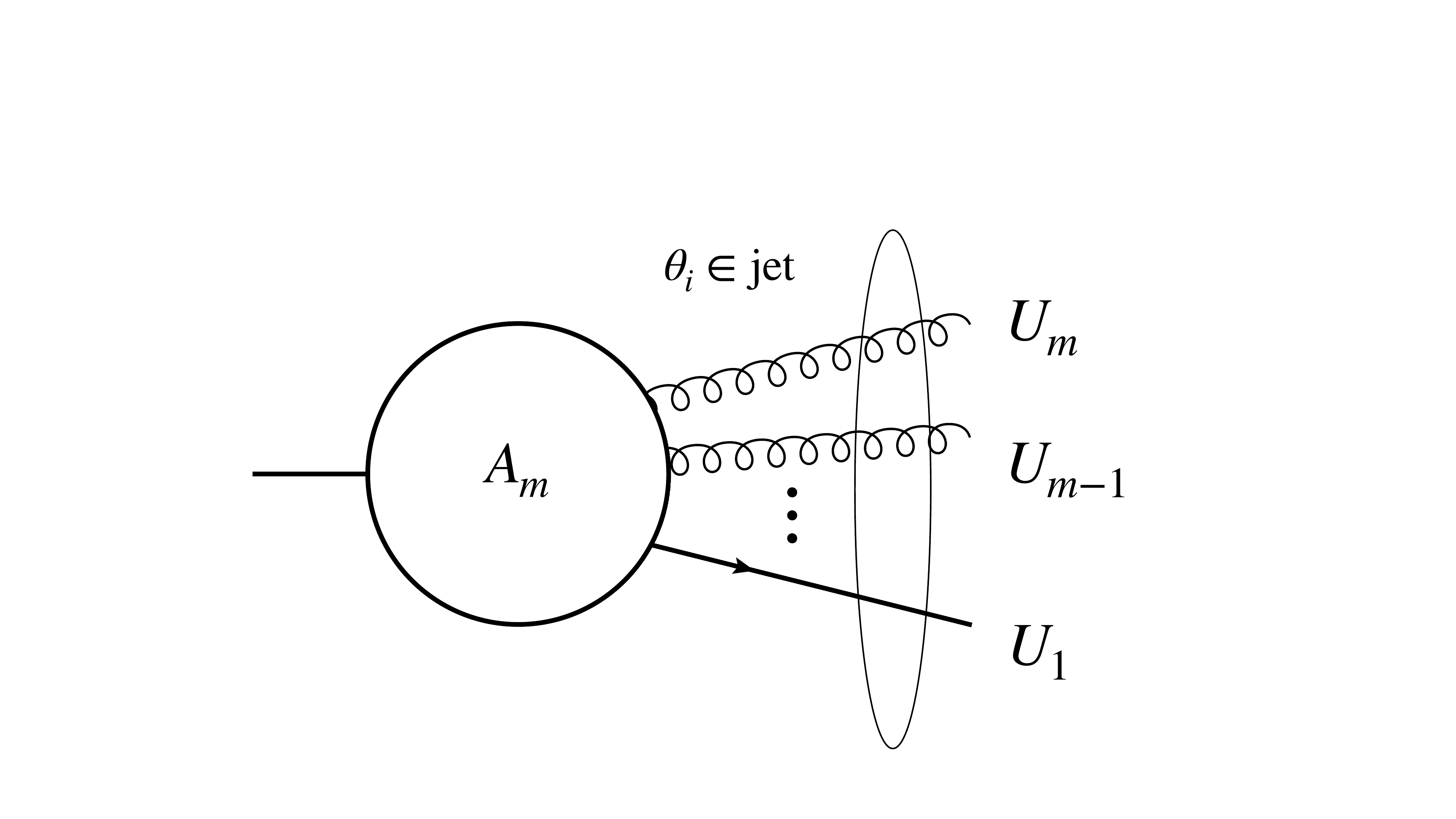}
\caption{Representation of the factorization \eqn{eq:cs-factorization} at the amplitude level, with $|A_m|^2 = C^{(m)}$. In the threshold limit collinear modes are confined inside the jet region.  \label{fig:fact-theorem-cs}}
\end{figure}

Accordingly, the general factorization formula that resums logarithms of $N$ must incorporate an explicit summation over collinear modes inside the jet \cite{Mehtar-Tani:2024smp,Mehtar-Tani:2025xxd}
\begin{align} \label{eq:cs-factorization}
& J_N(z,\mu) \nn 
&=\int_0^\infty \rmd\omega \, \delta((1-z)E-\omega)\,\sum_{m=1}^\infty \prod_{i=1}^m \int\frac{\rmd \Omega_i}{4\pi} C^{(m)}_{\rm coll}(\{n_i\},\mu) S^{(m)}_{\rm soft}(\{n_i\},\omega,\mu) +\cO(1/N) \,,
\end{align}
where $m$ denotes the number of resolved collinear partons inside the jet and $\{n_i\}\equiv\{n_1,\cdots, n_m\}$ denote the $m$ collinear directions. 
The collinear-soft operators are defined in terms of Wilson-line correlators along the directions of the collinear partons,
\beq \label{eq:soft-function-m}
S^{(m)} (\epsilon,R) \equiv  \sum_{X} \,\Theta_{\rm alg}\, \delta\!\left(\epsilon - \bar n \cdot p_{\rm out} \right) 
\langle 0| U_m^\dag \cdots U_1^\dag U_0^\dag | X \rangle  
\langle X| U_0  U_{1}\cdots U_m | 0 \rangle \,.
\eeq
The factorization formula eq.\,\eqref{eq:cs-factorization} is schematically represented in figure~\ref{fig:fact-theorem-cs} at the amplitude level.
Here, the sum runs over all final states $X$ of soft radiation. The operator $\Theta_{\rm alg}$ implements the jet algorithm constraint, ensuring that only radiation outside the jet cone of size $R$ contributes to the measurement. The delta function enforces the measurement of the total energy flowing outside the jet,
\beq
\epsilon = \bar n \cdot p_{\rm out}\,,
\eeq
where $p_{\rm out}$ is the total momentum carried by emissions outside the jet, and $\bar n$ is a light-like vector conjugate to the jet direction $n$. We choose light-like vectors $n^\mu=(1,0,0,1)$ and $\bar n^\mu=(1,0,0,-1)$, satisfying
$n^2=\bar n^2=0$ and $n\cdot\bar n=2$.

The operators $U_i \equiv U(n_i)$ denote collinear-soft Wilson lines in the fundamental or the adjoint representations, defined along the light-like directions $n_i$ of the collinear partons,
\beq\label{eq:WL}
U(n_i) = \mathcal{P} \exp\!\left[ ig \int_0^\infty \rmd s\, n_i \cdot A^a(s n_i)\, T^a \right]\,,
\eeq
where $\mathcal{P}$ denotes path ordering and $T^a$ are the color generators. The Wilson lines $U_{1},\dots,U_m$ correspond to the resolved collinear partons inside the jet. The Wilson line $U_0(\bar n)$ is an anti-collinear Wilson line along the $\bar n$ direction, which is required to ensure gauge invariance of the operator by completing the color structure of the hard scattering. As written, $S^{(m)}$ is a matrix in the color space of the $m+1$ partons, projected onto an overall singlet configuration by the contraction of color indices. Physically, this object encodes the coherent soft radiation emitted by a system of multiple color charges propagating along fixed directions. Despite its apparent complexity, the structure simplifies considerably in the large-$N_c$ limit, where the color algebra reduces to dipole interactions. A detailed discussion of such operators and their renormalization group evolution in vacuum can be found in Refs.~\cite{Becher:2016mmh,Caron-Huot:2015bja}.

Although we employ a common factorization scale $\mu$ throughout, it is important to emphasize the emergence of a new class of divergences for $m>2$, corresponding to configurations beyond single-parton energy loss. In particular, for $m>2$ the collinear matching coefficient $C^{(m)}$ develop an infrared (IR) divergence cancelled by UV divergences in $S^{(m)}$. In the absence of additional modes, this divergence would be cancelled by the corresponding virtual contribution. However, such a cancellation is incomplete once the jet is resolved by collinear-soft radiation. The origin of this mismatch lies in the differing structure of real and virtual contributions: $C^{(2)}$ describes configurations with two radiating partons (e.g., a quark and a gluon for an initial quark), whereas the virtual contribution encoded in $C^{(1)}$ with $\theta < R$ involves only a single emitter. This observation is at the core of the BMS equation that resums nonglobal lorarithms \cite{Banfi:2002hw} as well as the non-linear DGLAP evolution equation in the context of jet quenching \cite{Mehtar-Tani:2017ypq}.

As a result, real and virtual contributions do not cancel locally within the in-jet phase space once collinear-soft emissions are resolved. In the absence of the collinear-soft functions $S^{(1)}$ and $S^{(2)}$, unitarity would instead require, at order $\alpha_s$,
\beq
\int_{\theta < R} \rmd{\rm PS} \left( C^{(2)} + C^{(1)} \right) = 0\,,
\eeq
thereby ensuring the cancellation of infrared singularities in the unresolved limit. When the inner strucutre of jet are resolved collinear-soft gluons the cancellation is ensured via BMS evolution equations. 

In the Laplace representation, the jet cross section can be expressed as
\beq 
\frac{\rmd \sigma_{\rm jet}}{\rmd p_T} = H(p_T,\mu) \, C(\mu)\, \otimes \tilde S(\nu=N/p_T,\mu)+\mathcal{O}(1/N)+\mathcal{O}(R^2)\,,
\eeq
where we have used the $\otimes$ symbol as a shorthand for the matrix convolution over the $m$ collinear directions. Renormalization group (RG) consistency requires that the hard anomalous dimension be given by the sum of the soft and collinear contributions,
\beq \label{eq:rg-h-c-s}
\gamma_H(\mu) \equiv \mu \frac{\rmd}{\rmd \mu} \ln H(\mu)
= - \mu \frac{\rmd}{\rmd \mu} \ln C(\mu)
  - \mu \frac{\rmd}{\rmd \mu} \ln S(\mu)
= -\gamma_C(\mu) - \gamma_S(\mu)\,.
\eeq
Note that $\gamma_H$ is insensitive to soft-collinear scales and therefore does not probe the non-linear dynamics associated with the jet substructure, which is instead governed by a BMS-type renormalization group equation. More precisely, the collinear matching coefficient and the soft function depend on an additional factorization scale, such that $C \equiv C(\mu,\mu_{\rm cs})$ and $S \equiv S(\mu,\mu_{\rm cs})$. This is a consequence of the aforementioned new uncancelled divergences once collinear-soft emissions gets resolved. Thus, in addition to eq.\,\eqref{eq:rg-h-c-s}, RG consistency then requires
\beq \label{eq:rg-cs}
\mu_{\rm cs} \frac{\rmd}{\rmd \mu_{\rm cs}} \ln C(\mu,\mu_{\rm cs})
= - \mu_{\rm cs} \frac{\rmd}{\rmd \mu_{\rm cs}} \ln S(\mu,\mu_{\rm cs})\,,
\eeq
ensuring the cancellation of the dependence on the intermediate soft-collinear scale $\mu_{\rm cs}$. While one may set $\mu = \mu_{\rm cs}$ in practical implementations, such a choice obscures the distinct physical roles encoded in the different factors.

\paragraph{Double logarithmic approximation.} If we restrict ourselves to the double logarithmic approximation, in which the soft function $S$ describes soft and collinear radiation emitted by a single color charge. At leading order, one simply has
\begin{align}\label{eq:S-lo}
S_{\rm LO}(\epsilon) = \delta(\epsilon)\,, \qquad \tilde S_{\rm LO}(\nu) = 1\,.
\end{align}
At next-to-leading order in the double logarithmic approximation (DLA), the soft function takes the form
\begin{align}\label{eq:S-nlo}
S_{\rm DLA}(\epsilon) = \frac{\alpha_sC_F}{\pi} \int_R^1 \frac{\rmd \theta}{\theta}\int_0^{p_T}\frac{\rmd \omega}{\omega }\left[\delta(\epsilon-\omega)-\delta(\epsilon)\right] \,.
\end{align}
Its Laplace transform is given by
\beq 
\tilde S_{\rm DLA}(\nu) =  \frac{\alpha_sC_F}{\pi} \int_R^1 \frac{\rmd \theta}{\theta} \int_0^{p_T}\frac{\rmd \omega}{\omega } \left(\rme^{-\nu \omega}-1 \right)
\approx -  \frac{\alpha_sC_F}{\pi} \ln\frac{1}{R} \left[\ln (p_T \nu)+\gamma_E\right]\,.
\eeq
This expression makes explicit the phase space for soft radiation and, in particular, the origin of the leading logarithms $\ln R$ and $\ln(p_T \nu) \sim \ln N$, thereby connecting with the Mellin-space result in eq.\,\eqref{eq:nlo-mellin}. The dominant contribution arises from virtual soft gluons with energies in the interval
\beq
p_T \gg \omega \gg \nu^{-1} \sim \frac{p_T}{N}\,.
\eeq

To conclude, we emphasize that real emissions are strongly suppressed above the soft scale $p_T/N$. Consequently, the energy-loss dynamics is governed by Sudakov suppression, with real soft radiation contributing predominantly below this scale. In the next section we will compute the one loop correction to the total charge (single parton) that include the collinear and collinear-soft functions. 

\section{The collinear and collinear-soft functions at one-loop}
\label{sec:c-coft-function}
The goal of this section is twofold. First, we compute the one-loop anomalous dimension associated with a single collinear mode inside the jet. Second, we emphasize that this contribution is qualitatively distinct from higher-multiplicity configurations with $m>2$, where additional divergences arise and the dynamics becomes governed by non-linear evolution equations. At one loop, the evolution can be interpreted in terms of a single emitting parton. In contrast, at higher orders, the relevant degrees of freedom involve multiple correlated emitters, and the dynamics is more naturally described in terms of the total color charge of the jet. This behavior is expected, since soft radiation at large angles does not resolve the internal structure of the jet and is therefore sensitive only to its total charge, to all orders in perturbation theory.

\subsection{The total charge contribution}
\label{sub:total-charge-discussion}

Our strategy is then to factorize the contribution associated with the total color charge, which depends on the ultraviolet (UV) scale $\mu$, from the remaining dynamics, which encodes color-singlet evolution. Concretely, we rewrite the jet function as
\beq
J(\nu,\mu) = F_{\rm tot}(\mu,\nu) \sum_{m=1}^{\infty} \bar C^{(m)}(\mu_{\rm cs}) \otimes \bar S^{(m)}(\mu_{\rm cs})\Big|_{\rm sing}\,,\label{eq:total-charge-def}
\eeq
where
\beq
F_{\rm tot}(\mu,\nu) = C_{\rm tot}(\mu)\, S_{\rm tot}(\mu,\nu)
\eeq
represents the total-charge contribution. Here, $C_{\rm tot}(\mu)$ is the collinear factor associated with the parent parton, while $S_{\rm tot}(\mu,\nu)$ is the corresponding soft function describing radiation emitted from the parent direction $n$. 

This factorization cleanly separates the UV divergences, which are dominated by large-angle radiation and are entirely captured by $F_{\rm tot}$, from the remaining infrared-sensitive contributions. The latter arise in the coefficients $\bar C^{(m)}$ but are systematically canceled by the corresponding soft functions $\bar S^{(m)}$ when projected onto the color-singlet configuration. Note that as a result we have now an explicit decoupling of the $\mu$ and $\mu_{\rm cs}$. Although not strictly required, this procedure is physically well motivated and provides a transparent interpretation of the underlying dynamics.
\subsection{The collinear-soft function at one-loop}\label{sec:one-loop-soft-function}

In order to derive the renormalization group equations that describe the scale dependence of the soft-collinear and collinear factors, we shall now return to the calculation of the one-loop soft function whose operator definition reads
\beq \label{eq:soft-function-1}
&&S^{(1)} (\epsilon,R) \equiv \frac{1}{N_c} {\rm tr}_c \sum_{X}   \,\Theta_{\rm alg}\, \delta(\epsilon - \bar n \cdot p_{\rm out} ) \,  \langle 0|  U_1^\dag(n)U_0^\dag(\bar n) | X \rangle  \langle X| U_0(\bar n) U_{1}(n)| 0 \rangle  \,.
 \eeq
As is clear from this definition, the soft-collinear factor (and, likewise, the collinear matching coefficient) depends on the jet definition used to reconstruct the jets. In the following, we employ jet algorithms from the generalized $k_t$ family~\cite{Salam:2010nqg,Cacciari:2011ma} (anti-$k_t$~\cite{Cacciari:2008gp}, C/A, $k_t$~\cite{Dokshitzer:1997in,Wobisch:1998wt}, etc.), which are all equivalent at the one-loop order and in the small $R$ limit considered here~\cite{Ellis:2010rwa,Hornig:2016ahz,Kang:2016mcy,Marzani:2019hun}. Similar results can also be obtained for cone-type jet algorithms~\cite{Salam:2007xv,Kang:2017mda}.

\subsubsection{The out of cone contribution $\theta> R$ }
\begin{figure}[t]
\centering
\includegraphics[width=10cm]{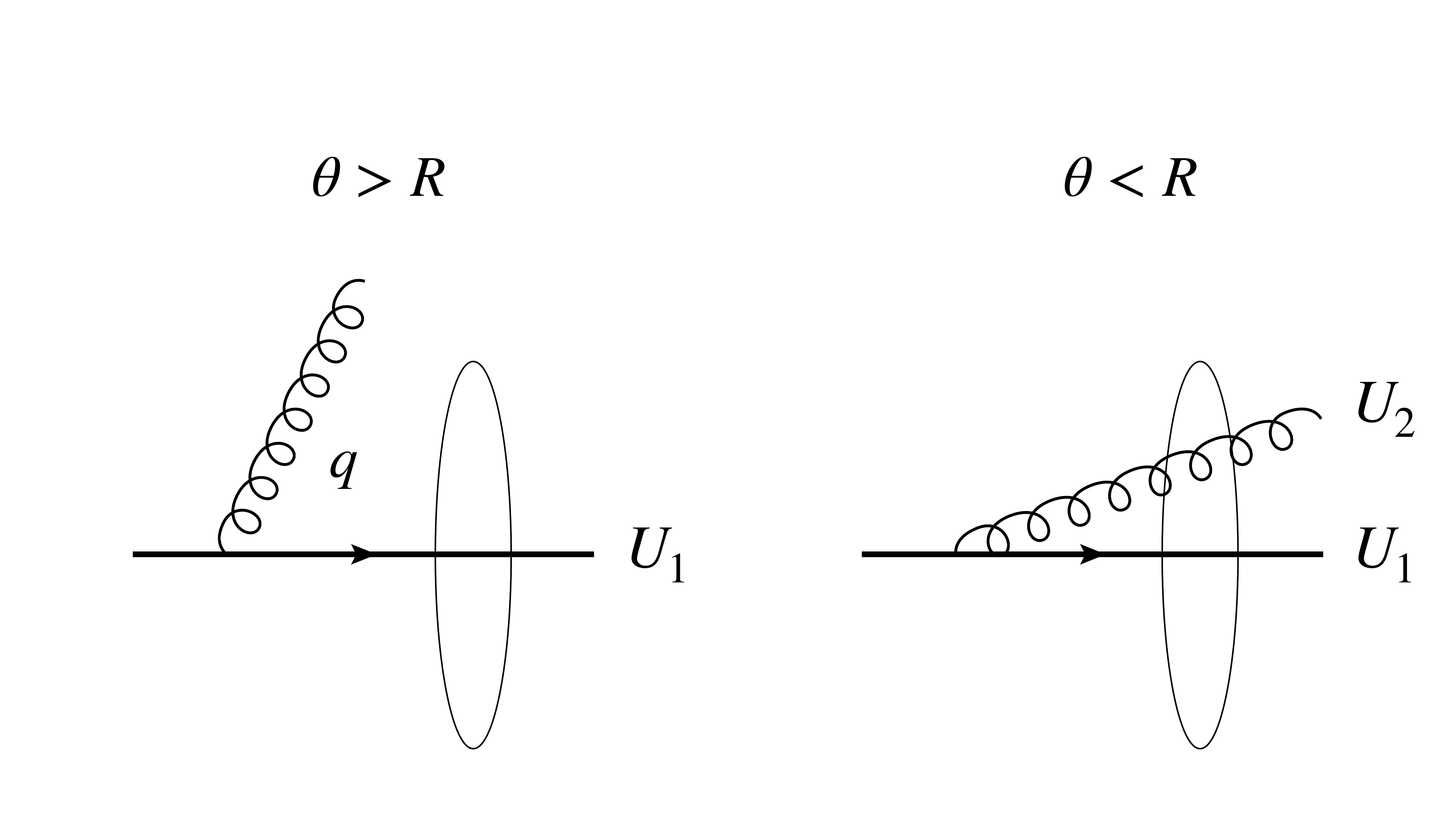}
\caption{Depiction of the out-of-cone (left) and in-cone (right) soft radiation. The subsequent evolution of the former cancels by unitarity, whereas the latter contributes to the observable through the acquisition of a Wilson-line phase, $U_2$.  \label{fig:S1-real}}
\end{figure}

\begin{figure}[t]
\centering
\includegraphics[width=11cm]{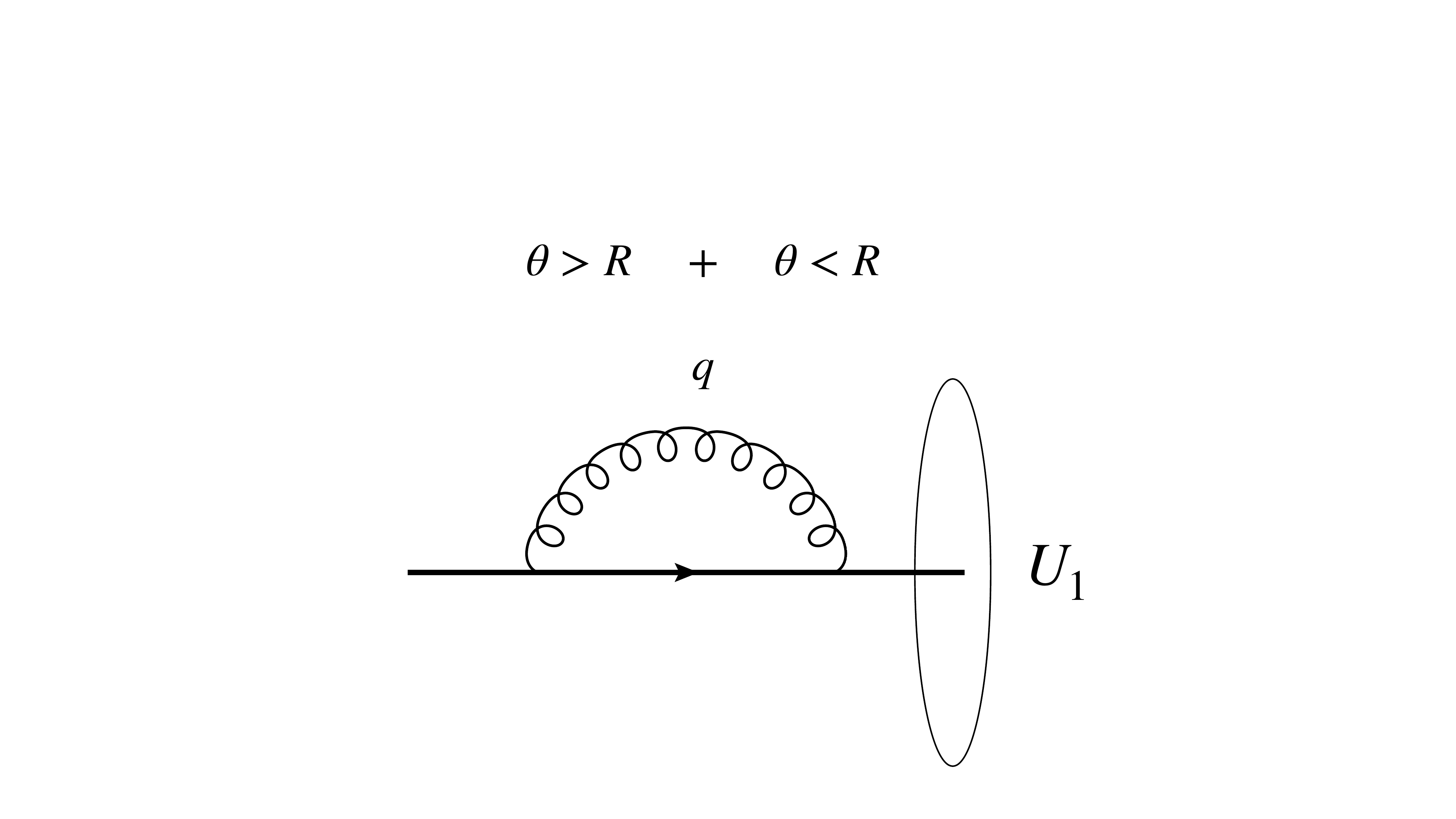}
\caption{The virtual correction to the soft function when $q^+ \ll E$. The in-cone and out-of-cone regions are split (the sum vanishes in dimensional regularization): the out-of-cone contribution combines with the corresponding real emission yielding \eqn{eq:s-one-loop} and the in-cone contribution is included in \eqn{eq:s-one-loop-in}. \label{fig:S1-virtual}}
\end{figure}

At one-loop, we have for the out of cone contribution 
\begin{align}\label{eq:s-one-loop}
& S^{(1)}_{\rm out}(\mu,k^+)= \delta(k^+) \nn &+ \frac{4g^2 C_F}{2(2\pi)} \left(\frac{\rme^{\gamma_E}\mu^2}{4\pi}\right)^{\epsilon} \int \frac{\rmd q^+}{q^+} \int \frac{\rmd^{2-2\epsilon}  q_\perp}{(2\pi)^{2-2\epsilon} q_\perp^2} \left[\delta(q^+-k^+)-\delta(k^+)\right] \Theta\left( |q_\perp|-\frac{R q^+}{2}\right)\,.
\end{align} 
The first (real) term inside the square bracket corresponds to the left diagram in figure~\ref{fig:S1-real} while the second term refers to the out-of-cone contribution of the virtual diagram depicted in figure~\ref{fig:S1-virtual}. Although one may compute this $k^+$ distribution directly, 
as we have alluded to above it is more convenient to switch to Laplace space, where the leading logarithms are more transparent. Hence, we can readily write
\begin{align}\label{eq:s-one-loop-2}
S^{(1)}_{\rm out}(\mu,\nu) &\simeq 1+ \frac{4g^2 C_F}{2(2\pi)} \left(\frac{\rme^{\gamma_E}\mu^2}{4\pi}\right)^{\epsilon}\int_0^{+\infty} \frac{\rmd q^+}{q^+}\left(\rme^{-q^+ \nu}-1\right)  \int_{|q_\perp|>\frac{R q^+}{2}} \frac{\rmd^{2-2\epsilon}  q_\perp}{(2\pi)^{2-2\epsilon} q_\perp^2}\,,\\
& = 1+ \frac{g^2 C_F}{\pi} \left(\frac{\rme^{\gamma_E}\mu^2}{\pi R^2}\right)^{\epsilon}\int_0^{+\infty} \frac{\rmd q^+}{(q^+)^{1+2\epsilon}}\left(\rme^{-q^+ \nu}-1\right)  \int^{+\infty}_{1} \frac{\rmd^{2-2\epsilon}  \btheta_\perp}{(2\pi)^{2-2\epsilon} \btheta^2}\,.
\end{align} 
Using the integrals in appendix~\ref{app:dimreg-integrals}, the final result reads
\beq 
S^{(1)}_{\rm out}(\mu,\nu) \simeq 1- \frac{\alpha_s C_F}{ 2 \pi}  \left[ \frac{1}{\epsilon^2}+ \frac{1}{\epsilon} L_{cs} + \frac{1}{2}\left( L_{cs}^2+\frac{\pi^2}{2}\right)\right]+\cO(\epsilon)\,,\label{eq:S1-outcone}
\eeq
where the collinear-soft logarithm is defined as 
\beq 
L_{cs} \equiv \ln \left( \frac{4e^{2\gamma_E}\mu^2 \nu^2 }{ R^2 }\right)\,.
\eeq
Recall that this is only the out contribution and one still needs to evaluate the in contribution that will involve non-linear terms with 4 Wilson lines. 
\subsubsection{The in of cone contribution $\theta< R$ }
The in-cone contribution is a sum of the real gluon radiation (right diagram in figure~\ref{fig:S1-real}) and the virtual part (cf.~figure~\ref{fig:S1-virtual}) that complements the one that was considered in the previous subsection when we computed the out of cone radiation. The real term depends on the soft-collinear operator $S^{(2)}$ in so far as the two prong structure of the real emission gets resolved. Combining the real and virtual terms, we then obtain
\begin{align}
S^{(1)}_{\rm in}(\mu,\nu) = \frac{g^2 C_F}{\pi} \left(\frac{\rme^{\gamma_E}\mu^2}{4\pi}\right)^{\epsilon}&\int \frac{\rmd q^+}{q^+} \int \frac{\rmd^{2-2\epsilon} \bq}{(2\pi)^{2-2\epsilon}\bq^2} \Theta_{\rm alg}(|\bq|< Rq^+/2)\nonumber\\
&\times\left[S_{12}^{(2)}(\nu) - S^{(1)}(\nu)\right]\,,
\end{align}
where we have used the simplified notation $S^{(2)}(\btheta_1,\btheta_2) \equiv S^{(2)}_{12}$.
In terms of angles, this equation reads
\begin{align} \label{eq:s-one-loop-in}
S^{(1)}_{\rm in}(\mu,\nu) &= \frac{\alpha_s C_F}{\pi^2}\left(\pi \rme^{\gamma_E}\mu^2\right)^{\epsilon}\int^{+\infty}_{\nu^{-1}}\frac{\rmd q^+}{(q^+)^{1+2\epsilon}} \int_0^{R} \frac{\rmd^{2-2\epsilon} \btheta_{12}}{\btheta_{12}^2} \left[S_{12}^{(2)}(\nu) - S^{(1)}(\nu)\right] \,,\\
&=\frac{\alpha_sC_F}{2\pi}\left[\frac{1}{\epsilon}+\ln\left(\pi e^{\gamma_E}\mu^2\nu^2\right)\right]\int_0^{R} \frac{\rmd^{2} \btheta_{12}}{\btheta_{12}^2} \left[S_{12}^{(2)}(\nu) - S^{(1)}(\nu)\right]\,.\label{eq:S1-incone}
\end{align}
The sum of eq.\,\eqref{eq:S1-incone} and eq.\,\eqref{eq:S1-outcone} provides the complete one-loop result for the soft-collinear operator $S_1^{(1)}$.

\subsection{The collinear function at one-loop}
We now turn to the collinear function $C$. In the threshold limit, the collinear function is proportional to $\delta(1-z)$ in physical space, and therefore to 1 in Laplace space. The coefficient of the delta function comes from (i) real collinear emission inside the jet, which exactly cancels against virtual corrections in the same phase space, (ii) virtual corrections in the out-jet phase space. Therefore, at one loop, the collinear function is obtained from the virtual corrections to a collinear splitting with the constraint $\theta>R$, namely
\begin{align}
C^{(1)} = 1-\frac{g^2}{2\pi}\left(\frac{\rme^{\gamma_E} \mu^2}{4\pi}\right)^\epsilon \int_0^1 \rmd z P(z,\epsilon) \int_0^{+\infty}\frac{\rmd^{2-2\epsilon} \bq }{(2\pi)^{2-2\epsilon}\bq^{2}} \Theta(|\bq|>R z(1-z) E/2)\,,
\end{align}
where $z=q^+/E$ and the splitting $qq$ function in $4-2\epsilon$ dimensions writes
\beq 
P(z,\epsilon) = C_F \left[\frac{1+z^2}{1-z} -\epsilon(1-z)\right]\,.
\eeq
The overall minus sign in $C^{(1)}$ comes from the fact that we are considering a virtual correction.
The $\bq$ integral reads   
\beq
C^{(1)} =1 -\frac{g^2}{8 \pi^2 \, \epsilon \,\Gamma(1-\epsilon) } \left(\frac{4 \rme^{\gamma_E} \mu^2}{E^2R^2}\right)^\epsilon \int_0^1 \frac{\rmd z}{[z(1-z)]^{2\epsilon}}  P(z,\epsilon) \,.
\eeq
Integrating over $z$ then expanding in powers of $\epsilon$, 
\begin{align}
C^{(1)}
&=1+
\frac{g^2 C_F}{8\pi^2}
\frac{1}{\epsilon\,\Gamma(1-\epsilon)}
\left(
\frac{4e^{\gamma_E}\mu^2}{E^2R^2}
\right)^\epsilon
\left[
\frac{1}{\epsilon}
+\frac{3}{2}
+\epsilon\left(
\frac{13}{2}-\frac{2\pi^2}{3}
\right)
+O(\epsilon^2)
\right]\,.
\end{align}
Then, 
\begin{align}
C^{(1)}
&=1-
\frac{\alpha_s C_F}{2\pi}
\left[
\frac{1}{\epsilon^2}
+\frac{1}{\epsilon}\left(L_c+\frac{3}{2}\right)
+\frac{1}{2}L_c^2
+\frac{3}{2}L_c
+\frac{13}{2}
-\frac{3\pi^2}{4}
+O(\epsilon)
\right],\label{eq:C1-oneloop-final}
\end{align}
where the collinear logarithm $L_c$ is defined as
\beq
L_c = \ln\!\left(\frac{4\mu^2}{E^2R^2}\right).
\eeq
One recovers the standard result for the integrated jet functions, previously obtained in~\cite{Ellis:2010rwa,Cheung:2009sg,Chay:2015ila}.
\section{Evolution equations and the large $N_c$ limit}
\label{sec:BMS-evol}
\subsection{Evolution equation for the one body distribution}

The evolution equation for the collinear-soft-function can be obtained using standard renormalization techniques. Before doing so, we note that we have derived the one-loop soft function for an off-centered collinear quark propagating at an angle $\theta_1$ with respect to the jet axis. Although this dependence does not play a role at one-loop order, it becomes essential at higher orders in the presence of multiple collinear partons. The key difference arises in the structure of the angular integration:

\beq
 \int_{|\btheta|>R}
\frac{\rmd^{2-2\epsilon}\btheta}{(2\pi)^{2-2\epsilon}}
\frac{1}{(\btheta-\btheta_1)^2}=
\frac{(4\pi)^\epsilon}{4\pi\,\Gamma(1-\epsilon)}
\left[
\frac{1}{\epsilon}
-
\ln\left(1-\frac{\btheta_1^2}{R^2}\right)
+\mathcal{O}(\epsilon)
\right].
\eeq
Therefore, combining the in and out one loop terms including this dependence on $\boldsymbol{\theta}_1$, we can readily obtain the divergent part of the one-loop correction to $S^{(1)}$ 
\begin{align}
    S_1^{(1)}(\mu,\nu)&=S^{(1)}_{\rm out}(\mu,\nu)+S^{(1)}_{\rm in}(\mu,\nu)\\
    &=1+\frac{\alpha_sC_F}{2\pi}\left\{-\frac{1}{\epsilon^2}-\frac{1}{\epsilon}\ln\left(\frac{4e^{2\gamma_E}\mu^2\nu^2}{R^2}\right)+\frac{1}{\epsilon}\ln\left(1-\frac{\boldsymbol{\theta}_1^2}{R^2}\right)\right.\nonumber\\
    &\left.+\frac{1}{\epsilon}\int_0^{R} \frac{\rmd^{2} \btheta_{12}}{\btheta_{12}^2} \left[S_{12}^{(2)}(\nu) - S_1^{(1)}(\nu)\right]+\textrm{finite}\right\}+\mathcal{O}(\alpha_s^2)\,.
\end{align}
with the shorthand notation $S^{(1)}(\btheta_1) \equiv S_1^{(1)}$. Strictly speaking, as noted in~\cite{Becher:2015hka,Becher:2016mmh}, the renormalization is not multiplicative as it involved the mixing with the operator $S^{(2)}$. From the pole structure, one reads the one loop evolution equation 
\begin{align}\label{eq:bms-full}
\frac{\rmd }{\rmd \ln \mu}S_1^{(1)}(\mu,\nu) &= \frac{\alpha_s C_F}{\pi}\left[-\ln \left( \frac{4e^{2\gamma_E}\mu^2 \nu^2 }{R^2 }\right)+\ln\left(1-\frac{\boldsymbol{\theta}_1^2}{R^2}\right)\right]S_1^{(1)}(\mu,\nu)\nonumber\\
&+\frac{\alpha_s C_F}{\pi^2} \int_0^{R} \frac{\rmd^{2} \btheta_{12}}{\btheta_{12}^2} \left[S_{12}^{(2)}(\mu,\nu) - S_1^{(1)}(\mu,\nu)\right]\,.
\end{align}
The first term in the square bracket of the first line resums to all orders corrections of the form $\alpha_s\ln(N/R)\ln(N)$ (since the evolution runs from $\mu=\nu\sim p_T/N$ to $\mu=p_T$), which were mentioned around eq.\,\eqref{eq:lnN-logs} in section~\ref{sec:scale}. Likewise, the evolution of equation for $C^{(1)}$ obtained from the pole structure of eq.\,\eqref{eq:C1-oneloop-final}, is given by
\begin{align}
    \frac{\rmd}{\rmd \ln \mu}C^{(1)}(\mu)&=\frac{\alpha_sC_F}{\pi}\left[\ln\left(\frac{4\mu^2}{E^2R^2}\right)+\frac{3}{2}\right]C^{(1)}(\mu)\,.
\end{align}
To relate the equation eq.\,\eqref{eq:bms-full} with the standard form of the BMS equation in the narrow jet limit, we now (i) isolate the total charge contribution discussed in subsection~\ref{sub:total-charge-discussion}, (ii) use the large $N_c$ limit to simplify the operator $S_{12}^{(2)}$. 
\subsection{Subtraction of the total charge}
As discussed around eq.\,\eqref{eq:J-refact-total}, the total charge soft-collinear factor does not depend on the unresolved internal substructure of the jet, and as such, it cannot depend on $\boldsymbol{\theta_1}$ and $\boldsymbol{\theta}_2$. Its evolution equation is therefore 
\begin{align}
    \frac{\rmd}{\rmd \ln \mu} S_{\rm tot}(\mu,\nu)=-\frac{\alpha_sC_F}{\pi}\ln\left(\frac{4e^{2\gamma_E}\mu^2\nu^2}{R^2}\right)S_{\rm tot}(\mu,\nu)\,,\label{eq:rg-Stot}
\end{align}
such that, using eq.\,\eqref{eq:total-charge-def}, the subtracted soft-collinear function $\bar S_1^{(1)}$ satisfies eq.\,\eqref{eq:bms-full} without the first term in the square bracket. eq.\,\eqref{eq:rg-Stot} is the Laplace space version of the renormalization group equation for the soft-collinear function derived in~\cite{Dai:2017dpc} where the intrajet soft-collinear modes are ignored. Likewise, the total charge collinear function is given at this order by $C_{\rm tot}(\mu)=C^{(1)}(\mu)$, again in agreement with~\cite{Dai:2017dpc}. In the notation of eq.\,\eqref{eq:total-charge-def}, we have therefore $\bar C^{(1)}=1+\mathcal{O}(\alpha_s^2)$ and $\bar S_1^{(1)}$ verifying
\begin{align}\label{eq:bms-full-S1sub}
\frac{\rmd }{\rmd \ln \mu}\bar S_1^{(1)}(\mu,\nu) &= \frac{\alpha_s C_F}{\pi}\ln\left(1-\frac{\boldsymbol{\theta}_1^2}{R^2}\right)\bar S_1^{(1)}(\mu,\nu)\nonumber\\
&+\frac{\alpha_s C_F}{\pi^2} \int_0^{R} \frac{\rmd^{2} \btheta_{12}}{\btheta_{12}^2} \left[S_{12}^{(2)}(\mu,\nu) - \bar S_1^{(1)}(\mu,\nu)\right]\,.
\end{align}
This  evolution equation is evidently not closed, since it involves an infinite hierarchy of coupled equations. To solve for $\bar S^{(1)}$, one requires the equation for $\bar S^{(2)}$, which itself depends on $\bar S^{(3)}$, and so on. However, in the large-$N_c$ limit, the gluon--quark parton energy-loss distribution factorizes into a quark distribution times a dipole (color-singlet) distribution:
\beq
S^{(2)}_{12} \simeq S_2\, S_{12} \, .
\eeq
Therefore, the color structure in \eqn{eq:bms-full-S1sub} becomes
\beq
S^{(2)}_{12} - S_1 \;\to\; S_{12} S_2 - S_1 \, .
\eeq
\subsection{The antenna evolution: the BMS equation}
Using the large $N_c$ limit, we then find
\beq \label{eq:bms-full-large-Nc}
\frac{\rmd }{\rmd \ln \mu}\bar S_1^{(1)}(\mu,\nu) &=& \frac{\alpha_s N_c}{2\pi}\ln \left(1-\frac{\btheta_1^2}{R^2}\right)\,\bar S_1^{(1)}(\mu,\nu)\nn
&+&\frac{\alpha_s N_c}{2\pi^2} \int_0^{R} \frac{\rmd^{2} \btheta_{12}}{\btheta_{12}^2} \left[\bar S_{12}(\mu,\nu)\bar S_2^{(1)}(\mu,\nu)- \bar S_1^{(1)}(\mu,\nu)\right]\,,
\eeq
which is nothing but the BMS equation in the narrow jet limit~(cf.~eq.\,(6.26) in~\cite{Banfi:2002hw}).

In the large-$N_c$ limit, the system closes through the evolution equation for the color-singlet dipole distribution, which is nothing but the BMS/BK equation:
\begin{align}
    \label{eq:bms-12}
\frac{\rmd \bar S_{12}}{\rmd \ln \mu} 
&=\frac{\alpha_s N_c}{2\pi}\ln\left[1+\frac{\btheta_{12}^2 R^2}{(R^2-\btheta_1^2)(R^2-\btheta_2^2)}\right] S_{12}
+\frac{\alpha_s N_c}{2\pi^2}
\int_0^{R} \rmd^{2}\btheta_{3}\,
\frac{\btheta_{12}^2}{\btheta_{13}^2 \btheta_{32}^2}
\left[
S_{13} S_{32} - S_{12}
\right] \, .
\end{align}
In vacuum, the boundary conditions for $S_{12}$ and $S_1$ are simply
\beq
S_{12}(\mu=\nu^{-1}) = S_{1}^{(1)}(\mu=\nu^{-1}) = 1 \, .
\eeq
From the perspective of the ultraviolet structure of the Wilson-line operators, this boundary condition is imposed at light-cone infinity. In the presence of a medium, the ultraviolet structure remains unchanged; the medium modifies only the boundary condition, as we discuss below. The resulting structure takes a similar form as high-energy factorization with non-linear QCD evolution equations. The BMS equation, which governs energy flow away from a jet in the vacuum, replaces the BK equation, which describes gluon saturation in the proton at high energy. Both equations are based on strong energy ordering of soft gluon emissions and are known to be dual to one another through a stereographic mapping. A crucial difference, however, is that the BMS equation resums soft radiation in the vacuum, whereas the BK equation encodes multiple interactions with a target proton state, thereby loosening the physics connection. In this work, we push this duality further by extending the BMS equation to include interactions with a QCD medium. This seemingly minimal extension has far-reaching consequences, as it enables the incorporation of the complex physics of jet energy loss within a unified and coherent theoretical framework.  

\section{Boundary condition for the in-medium soft function}
\label{sec:med-boundary}

The key assumption underlying this extension of our formalism from proton-proton collisions to heavy-ion collisions is the effective decoupling between the hard scattering process responsible for jet production and the strongly interacting plasma that constitutes the surrounding medium. More precisely, we assume that correlations between the initial state of the quark-gluon plasma, treated as a large thermal environment, and the short-distance dynamics generating the energetic jet are parametrically suppressed. As a consequence, the initial density matrix of the jet factorizes from the density matrix describing the medium, with any residual entanglement or correlation suppressed by inverse powers of the system size.
Within the open quantum system framework, the jet is treated as a localized subsystem propagating through and interacting with an external environment composed of the soft degrees of freedom of the quark--gluon plasma~\cite{Mehtar-Tani:2024smp,Mehtar-Tani:2025xxd}. Following its production in a hard scattering process, the jet experiences multiple soft final-state interactions with medium constituents. Because these interactions involve long-wavelength color exchanges, they do not resolve the microscopic structure of the hard process itself. Instead, they lead to collective effects such as color decoherence, transverse momentum broadening, and medium-induced radiation. The influence of the plasma on the jet evolution is therefore encoded in expectation values of Wilson-line operators evaluated in the medium state rather than in the vacuum. In this manner, the many-body and nonperturbative properties of the plasma enter through medium correlators of Wilson lines, which generalize the vacuum matrix elements familiar from proton--proton collisions and provide a gauge-invariant description of color transport through the medium.

\subsection{Separation of vacuum and medium time scales}
The matrix elements of the collinear-soft operators are evaluated in the medium state instead of the vacuum. For example, the single-parton energy-loss distribution, or equivalently the collinear-soft function, is modified according to
\beq \label{eq:soft-function-med}
\langle 0| U_0^\dag U_1^\dagger | X \rangle  
\langle X| U_1 U_0| 0 \rangle
\quad \longrightarrow \quad
\langle \med| U_0^\dag U_1^\dagger | X \rangle  
\langle X| U_1 U_0| \med \rangle \, .
\eeq
The quantum fluctuations encoded in this function remain soft relative to the hard scale $p_T$ characterizing the hard and collinear sectors. At this stage, these matrix elements may still be computed directly within full QCD. Alternatively, one may construct an additional effective field theory in which Glauber modes~\cite{Rothstein:2016bsq}, responsible for Coulombic scattering between jet constituents and plasma particles, are systematically integrated out \cite{Vaidya:2020cyi}. 

In this work, however, we adopt a different strategy. We first implement a Wilsonian separation between the collinear-soft modes, with scaling
\beq
k^+ \sim E_{\rm loss} \sim \frac{p_T}{N}\, ,
\eeq
and the medium soft modes,
\beq
k^+ \sim T\, ,
\eeq
which, in the center-of-mass frame of the jet--medium system, may be regarded as anti-collinear modes. The formalism employed is based on the background field method, where the separation of modes is imposed directly on the $+$ light-cone momentum component. This differs conceptually from the standard SCET construction~\cite{Bauer:2000yr,Bauer:2002nz,Bauer:2002aj,Bauer:2003mga,Bauer:2011uc}, in which the scaling of all momentum components follows from the on-shell condition and the associated power counting.

In vacuum, covariant gauges are often preferred because they preserve manifest Lorentz covariance. In the presence of a medium, however, it is generally more convenient to work in the physical light-cone gauge, which suppresses radiation from the anti-collinear sector --- namely, radiation emitted by the medium color sources --- and simultaneously reduces the Wilson line $U_0$ to the identity. Consequently, for the gauge choice
$\bar n \cdot A = A^+ = 0,$ one obtains $U_0 = 1$.

In the previous section, we discussed the ultraviolet behavior of the collinear-soft function, which is insensitive to the medium dynamics since it probes scales satisfying $\omega \gg p_T/N \sim E_{\rm loss}$. Medium-induced interactions occur only at the boundary of this phase space and therefore determine the initial condition for the RG evolution in \eqn{eq:bms-full}, in a manner analogous to the role of the McLerran--Venugopalan (MV) model as the initial condition for small-$x$ evolution.

We need therefore two initial distributions for the single quark energy loss distribution $S_1$ and the antenna (quark-antiquark) energy loss distribution $S_{12}$ that has been computed in the large medium approximation in \cite{Mehtar-Tani:2017ypq}
\beq
\bar S^{\rm med}_1\equiv S_1(\mu=\nu^{-1},\nu)\,
\qquad \text{and}\qquad
\bar S^{\rm med}_{12}\equiv S_{12}(\mu=\nu^{-1},\nu)\,,
\eeq
where $\nu \sim p_T/N$\,. Two complementary strategies may be employed to characterize these quantities. First, $S_1$ and $S_{12}$ may be treated as a genuinely nonperturbative object and parametrized phenomenologically, with its parameters constrained through Bayesian inference against experimental data (following the approach of~\cite{Qiu:2019sfj}). Alternatively, it may be computed analytically or numerically using perturbative techniques, under the assumption that the interactions with the quark-gluon plasma remain predominantly in the weak-coupling regime.

\subsection{Single-prong energy-loss distribution}
Returning to the analogy with small-$x$ evolution, recall that in the MV model the multiple scatterings are treated as quasi-instantaneous within the shockwave approximation relative to the quantum fluctuations resummed by the BK equation. In the present case, by contrast, the multiple interactions with the medium may be regarded as occurring over parametrically long timescales, effectively at $t\sim L \to +\infty$, where $L$ denotes the size of the medium.

\begin{figure}[t]
\centering
\includegraphics[width=12cm]{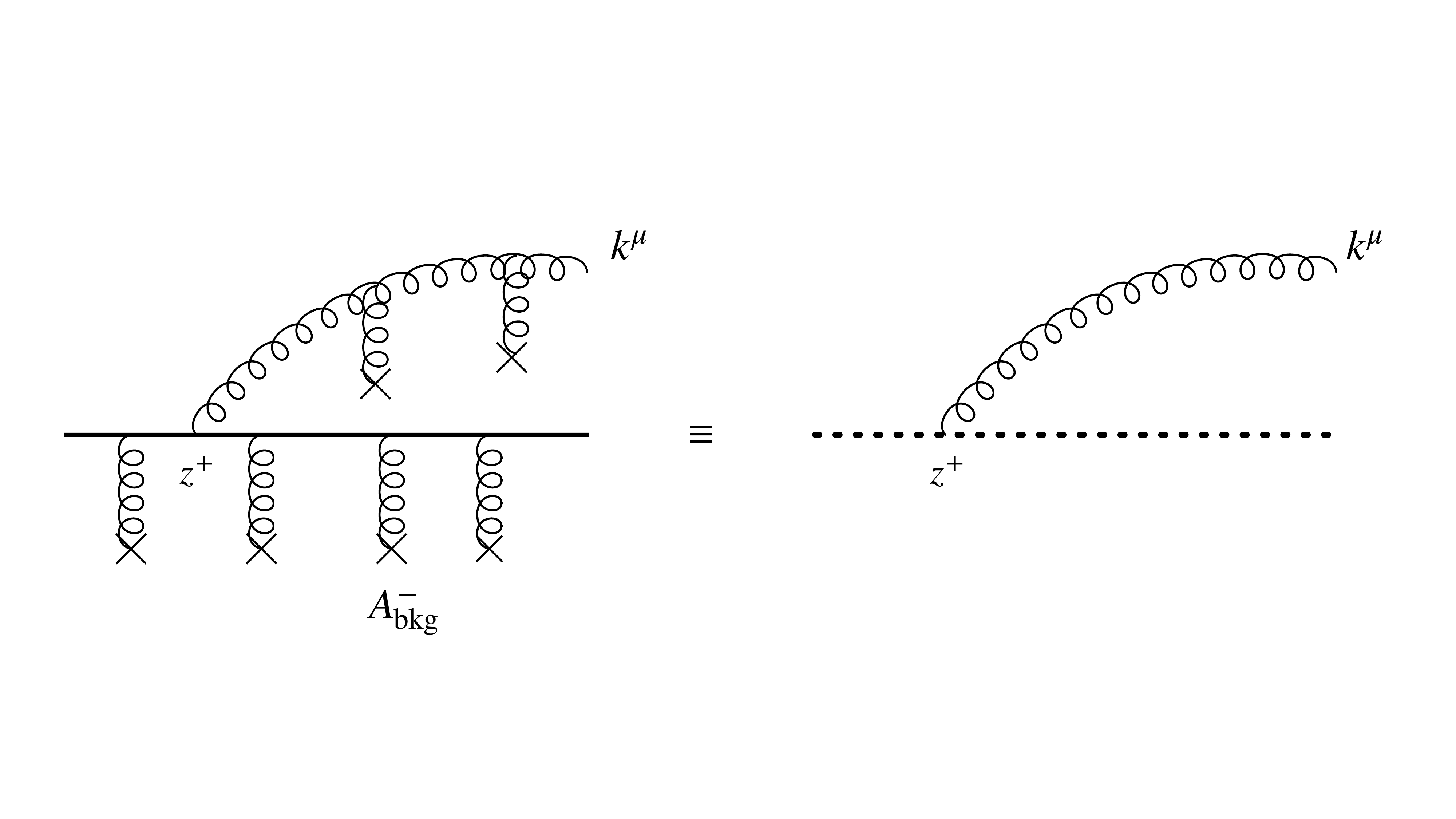}
\caption{Diagrammatic representation of the resummation of multiple interactions with the medium background field. An arbitrary number of insertions of $A^-_{\rm bkg}$ along the eikonal parton and the emitted gluon of momentum $k^\mu$ is absorbed into background-field Wilson lines and the corresponding dressed gluon propagator, represented by the dotted line. The emission occurs at light-cone time $z^+$.\label{fig:single-med-emission}}
\end{figure}

At leading order in perturbation theory, in the eikonal approximation, $S_1^{\rm med}$ is related to the medium-induced radiative spectrum denoted as $I_{\rm med}$ (cf. figure~\ref{fig:single-med-emission}): 
\beq 
\bar S^{\rm med}_1(\nu) =1 + \int_0^\infty \rmd q^+ \int \frac{\rmd^2 \bq}{(2\pi)^2}\frac{\rmd I_{\rm med}}{\rmd q^+ \rmd^2\bq } \, \left(\rme^{-\nu \omega} \Theta(|q_\perp|-Rq^+/2)-1\right)+\cO(\alpha_s^2)\,.
\eeq
Once again, the above expression exhibits an explicit ultraviolet divergence; however, choosing $\mu=\nu^{-1}$ ensures that the corresponding vacuum logarithms are minimized.

\begin{figure}[t]
\centering
\includegraphics[width=8cm]{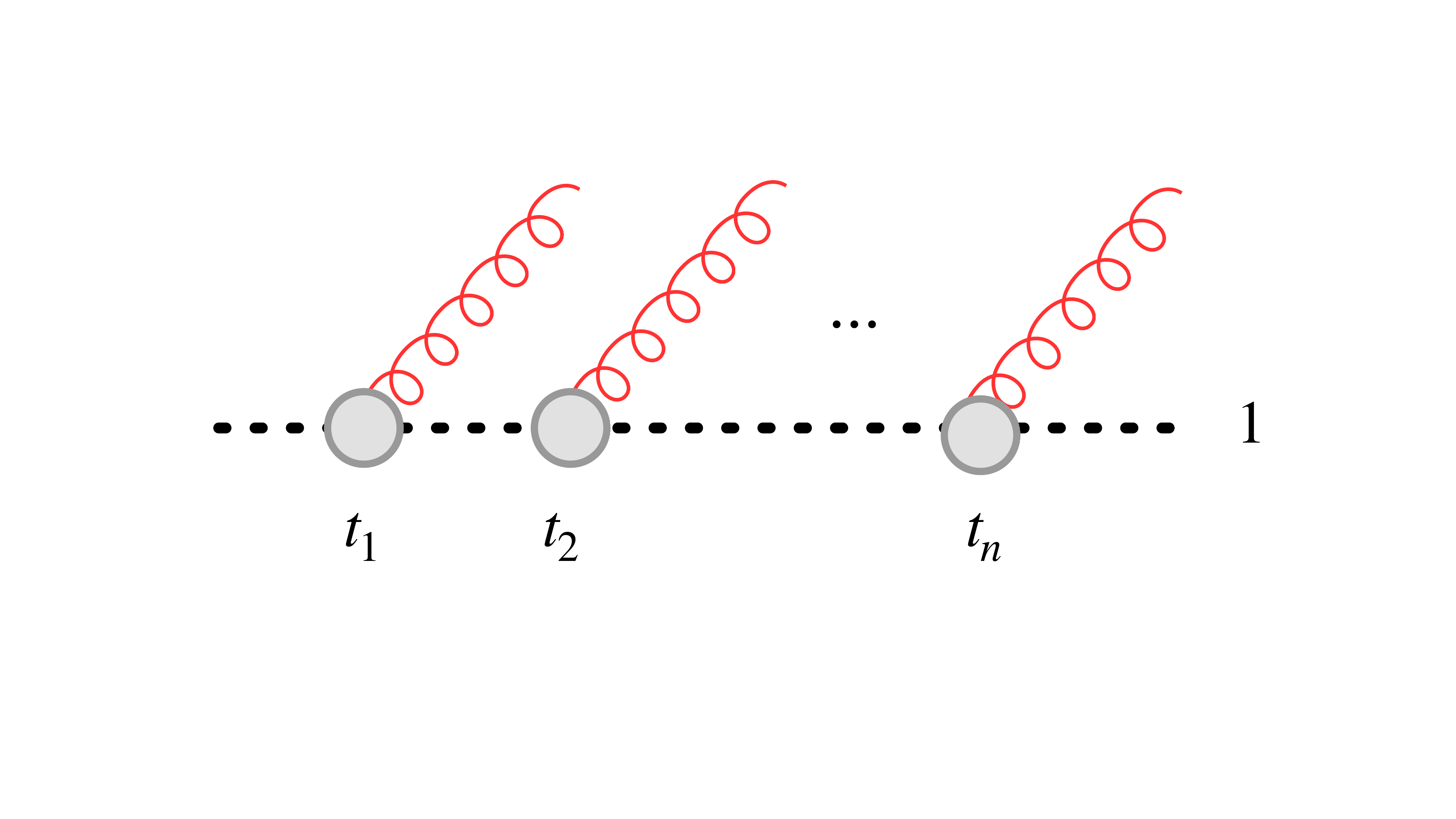}\includegraphics[width=8cm]{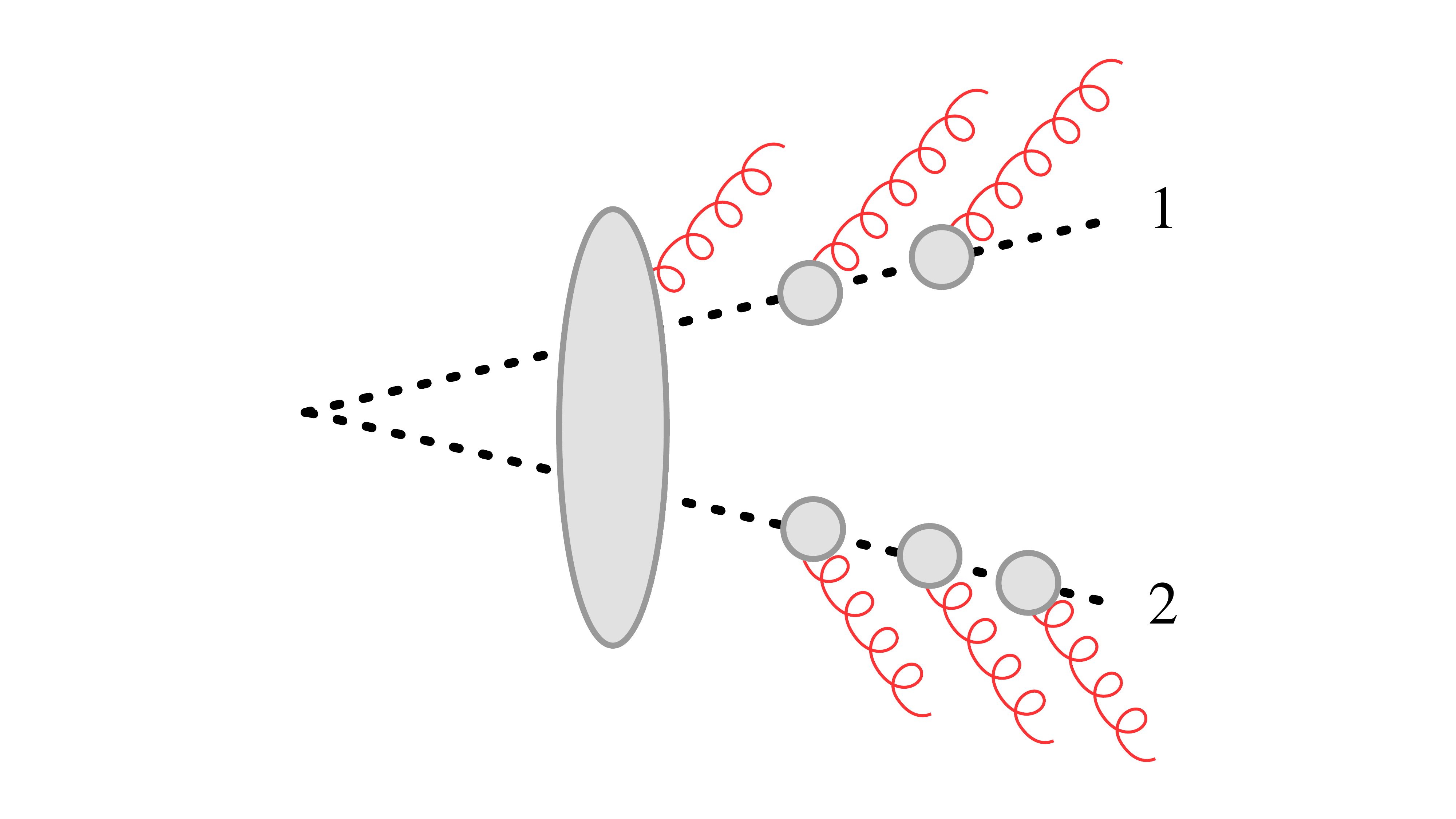}
\caption{Schematic representation of medium-induced radiation from a one-prong (left) and a two-prong (right) system. Successive emissions occur at times $t_1,\ldots,t_n$ along each resolved color charge. For the two-prong system, the medium resolves the daughter partons only after the decoherence time indicated by the shaded region, after which the two prongs radiate independently.\label{fig:evol-antenna}} 
\end{figure}

An interesting regime to consider is that of a large medium. To simplify the discussion, we neglect subsequent branchings of the primary emitted gluons. Although this approximation captures the relevant time-scale ordering between vacuum and medium-induced radiation, a complete treatment must include multiple branching processes. This generalization is straightforward and was studied in Ref.~\cite{Mehtar-Tani:2024mvl}. Since the medium interactions are localized within a finite correlation length $\ell$, one generically expects large enhancements proportional to $L/\ell \gg 1$, where $L$ denotes the medium size. Such enhancements require resummation and are naturally treated within a kinetic framework ($\omega\equiv q^+$) \cite{Jeon:2003gi,Baier:2001yt}:
\beq
\frac{\partial}{\partial t }\bar S^{\rm med}_1(\epsilon,t)
= \int_0^{\infty} \rmd \omega \,
\frac{\rmd \Gamma (\abar,\omega)}{\rmd \omega}
\left[
\bar S_1^{\rm med}(\epsilon+\omega,t )-\bar S_1^{\rm med}(\epsilon,t)
\right]\,.
\eeq
or in Laplace space
\beq
\frac{\partial}{\partial t }\bar S^{\rm med}_1(\nu,t)
= \int_0^{\infty} \rmd \omega \,
\frac{\rmd \Gamma (\abar,\omega)}{\rmd \omega}
\left(\rme^{-\nu \omega}-1\right)\bar S_1^{\rm med}(\nu,t )\,.
\eeq
An illustration of this equation is given in the left cartoon of Figure~\ref{fig:evol-antenna}.

Here, $\Gamma (\abar,\omega)= \rmd I_{\rm med}/\rmd t +\cO(\abar^2) $ denotes the energy-loss rate, which can in principle be computed order-by-order in perturbation theory. Determining $\Gamma (\abar,\omega) $ beyond leading order is, however, highly nontrivial (see the series of papers~\cite{Arnold:2015qya,Arnold:2016kek,Arnold:2016mth,Arnold:2016jnq,Arnold:2020uzm,Arnold:2022epx,Arnold:2022fku} for the NLO calculation). Even at the level of single-gluon emission, it is necessary to resum multiple scatterings occurring during the formation time. This problem has been studied extensively in the literature, and we refer the reader to the relevant references~\cite{Baier:1996kr,Baier:1996sk,Zakharov:1996fv,Zakharov:1997uu,Gyulassy:2000er,Wiedemann:2000za} for more comprehensive discussions.
In this work, our primary goal is to emphasize the renormalization-group and kinetic structures underlying jet evolution in the medium. We therefore restrict ourselves, whenever possible, to general and model-independent formulations. Nevertheless, for illustrative purposes, we will occasionally adopt simplifying assumptions that allow us to make more direct contact with the underlying medium dynamics.

Turning now to the medium-induced contribution, assuming that gluons are dominantly radiated out of the jet region, we have
\begin{align}
\tilde S_\med^{(1)}(\nu) &=  \int_0^L \rmd t \int_0^{p_T}\frac{\rmd \omega}{\omega } \Gamma(\omega,t)\left(1-\rme^{-\omega \nu} \right)\,,\label{eq:S1-BDMPS}\\
&\sim  \abar L \sqrt{\hat q } \int_{\nu^{-1}}^{p_T}\frac{\rmd \omega}{\omega^{3/2} } \sim \sqrt{\nu}, 
\end{align}
where the radiation rate reads \cite{Mehtar-Tani:2025rty}
\beq 
\Gamma(\omega,t) = \abar \sqrt{\frac{\hat q}{\omega}}\,,
\eeq
with $\bar\alpha_s=\alpha_s N_c/\pi$. Contrary to the vacuum radiation that is logarithmic the above integral is dominated by the energy scale $p_T/N$ due to the $\omega^{-3/2}$ fall off of the radiation spectrum. Medium-induced gluons have longer formation time than vacuum radiation and are emitted with constant rate throughout the medium of size $L$ contrary to vacuum bremsstrahlung that is tied to the hard vertex. Hence, in the leading log approximation and large medium length limit there is no overlap between the two emission mechanisms and as a result they factorize in Laplace space justifying the fact that the medium distribution sets the initial condition for vacuum evolution. 

The simplified expression eq.\,\eqref{eq:S1-BDMPS} is presented for illustrative purposes and does not entail any loss of generality. A more rigorous, field theoretical, definition for the radiative spectrum whose derivative w.r.t. to $L$ yields the rate $\Gamma$ is given by the Wilson-line correlator, assuming all the energy is radiated outside the jet for simplicity:  
\begin{align}\label{eq:operator-spectrum}
\omega \frac{\rmd I_{\rm med}}{\rmd\omega}
=
\frac{\alpha_s}{2N_c}
\int_{0}^{L} \rmd t'\,
\int_{0}^{t'} \rmd t\,
\left\langle
\operatorname{Tr}
\,U_1^{\dagger}(t',t)\,
\cW_{11} (t',t)
\right\rangle-\rm{vac.}
\end{align} 
where 
\begin{align} \label{eq:inter-matrix}
\mathcal{W}_{ij} &\equiv \mathcal{W}(\bx_j,t';\bx_i,t)\,,\\
&\equiv
\frac{1}{\omega^2}
\left(\bdel_x-i\omega \bn_j\right)\!\cdot\!
\left(\bdel_y+i\omega \bn_i\right)\,
\mathcal{G}(\bx,t';\by,t)\,
e^{-i\frac{\omega}{2}\bn_j^2 t'
+i\frac{\omega}{2}\bn_i^2 t}
\Big|_{\substack{\by=\bx_i(t)\\ \bx=\bx_j(t')}}\,.
\end{align} 
Here, $U_1$ and $\cG$ are the background field Wilson-line and the $2+1$ dimensional scalar propagator in the adjoint representation obeying the equations: 
\beq 
\left[\frac{\del}{\del x^+} -ig A^{-,a}_\bkg(x^+,\bx=\bn x^+)T^a\right] \, U_1(x^+,y^+) =0\,,
\eeq
and 
\beq 
\left[\frac{\del}{\del x^+} - \frac{\bdel^2_x}{2 k^+}-ig A^{-,a}_\bkg(x^+,\bx)T^a\right]\mathcal{G}(\bx,t';\by,t) = i\delta (\bx -\by)\delta (x^+ -y^+)\,,
\eeq
which obeys a Schrödinger equation describing the dynamics of a non-relativistic particle of mass $k^+$ in $2+1$ dimensions. eq.\,\eqref{eq:operator-spectrum} has been extensively studied and evaluated both in the harmonic-oscillator approximation, which describes the multiple-soft-scattering regime, and at leading order in the opacity expansion. For a recent review, see Ref.~\cite{Mehtar-Tani:2025rty} and references therein.

\subsection{Color decoherence of a two-prong system}

Perturbative expressions for the single-quark energy loss have long been known, and more recently analogous results have been derived for the energy loss of a quark--antiquark antenna~\cite{Mehtar-Tani:2017ypq}. Owing to an incomplete understanding of the underlying QCD structure, however, the latter --- whose full form emerges at next-to-leading order --- has so far not found concrete phenomenological applications. The framework developed here enables such applications for the first time.

To keep the discussion at the level of the interplay between vacuum and medium-induced dynamics, we shall assume that medium-induced radiation withing the jet cone do not contribute to the observable. Under these simplifications the initial condition for $\bar S_{12}$ obeys the following rate equation, where $t$ is the real time variable or can also be viewed as the system size $L$ \cite{Mehtar-Tani:2017ypq}:

\begin{figure}[t]
\centering
\includegraphics[width=12cm]{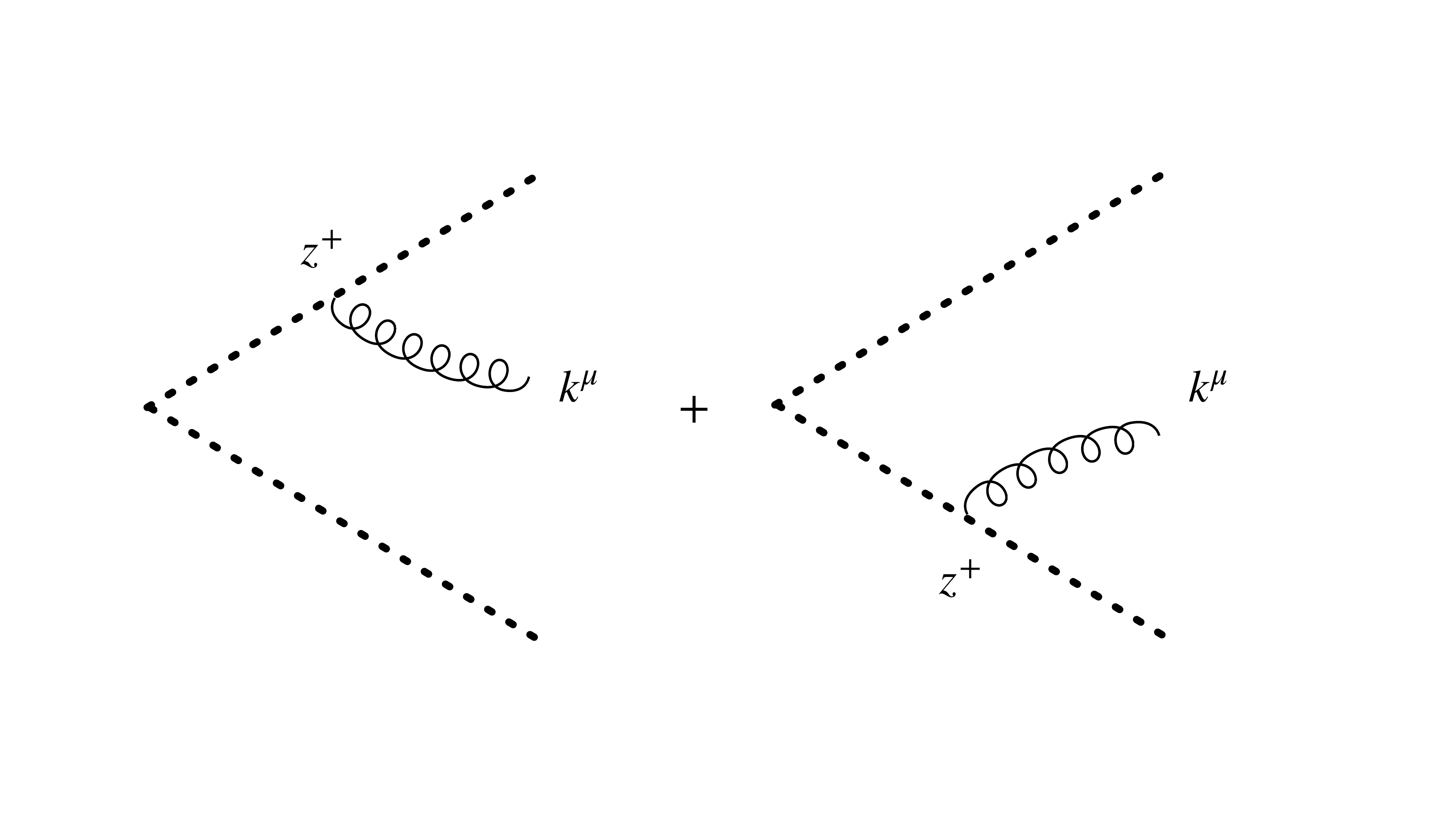}
\caption{Real-emission contributions from a color dipole. A gluon with momentum $k^\mu$ is emitted at light-cone time $z^+$ from either of the two eikonal legs. The sum of the two amplitudes encodes the interference responsible for color-coherence effects. Interactions with the background field are implicit.\label{fig:interference-graph}}
\end{figure}

\begin{align}
\bar S_{12}^{\rm med}(\epsilon,L)
&= \delta(\epsilon)
+ \int_0^{L} \rmd t \int_0^\infty \rmd\omega \,
\bigl[\Gamma_{11}(\omega,t)+\Gamma_{22}(\omega,t)\bigr]
\, \bar S_{12}^{\rm med}(\epsilon-\omega,t)
\nn
&\quad
+ \int_0^{L} \rmd t \,
\bigl[1-\Delta_{\mathrm{med}}(t)\bigr]
\int_0^\infty \rmd\omega \,
\bigl[\Gamma_{12}(\omega,t)+\Gamma_{21}(\omega,t)\bigr]
\, \delta(\epsilon-\omega) \, .
\end{align}

An illustration of this equation is given in the right cartoon of Figure~\ref{fig:evol-antenna}.
The interference rates introduced in~\cite{Mehtar-Tani:2017ypq} are built from the matrix element $\cW$ given in \eqn{eq:inter-matrix}, defined as 
\begin{align}\label{eq:operator-spectrum}
\omega \frac{\rmd \Gamma_{ji}}{\rmd\omega}
=
\frac{\alpha_s}{2N_c}
\int_{0}^{\infty} \rmd \tau\,
\left\langle
\operatorname{Tr}
\,U_i^{\dagger}(t+\tau,t)\,
\cW_{ij} (t+\tau,t) 
\right\rangle-\rm{vac.}
\end{align} 
and obtained from the interference between the two diagrams shown in figure~\ref{fig:interference-graph}.

In Laplace space the corresponding rate equation reads
\begin{align}
\frac{\partial}{\partial t}\,\bar S_{12}^{\rm med}(\nu,t)
=
\Gamma_{\rm {dir}}(\nu,t)\,\bar S_{12}^{\rm med}(\nu,t)
+
\Gamma_{\mathrm{int}}(\nu,t),
\end{align}
where
\begin{align}
\Gamma_{\mathrm{dir}}(\nu,t)
&=
\int_{0}^{\infty} d\omega\,
\Bigl(\Gamma_{11}(\omega,t)+\Gamma_{22}(\omega,t)\Bigr)
e^{-\nu\omega}\,,\\
\Gamma_{\mathrm{int}}(\nu,t)
&=
\Bigl[1-\Delta_{\mathrm{med}}(t)\Bigr]
\int_{0}^{\infty} d\omega\,
\Bigl(\Gamma_{12}(\omega,t)+\Gamma_{21}(\omega,t)\Bigr)
e^{-\nu\omega}.
\end{align}
The solution to the rate equation is
\begin{align}
\label{eq:main-result}
\bar S_{12}^{\rm med}(\nu,L) &=   \bar S_1^{\rm med}(\nu, L) \, \bar S_2^{\rm med}(\nu, L)\,  \nn 
& - 2 \int_0^L \rmd t  \, \bar S_1^{\rm med}(\nu, L-t) \, \bar S_2^{\rm med}(\nu, L-t)  \Big[ 1- \Delta_\med(t) \Big]  \, \Gamma (\omega, t)\,.
\end{align}
Here, $\bar S_{12}^{\rm med}(\nu,L)$ describes the independent energy loss of the antenna legs, and $\Delta_\text{med}(t)$ is the so-called decoherence parameter \cite{Mehtar-Tani:2011hma,Mehtar-Tani:2011vlz,Mehtar-Tani:2012mfa}
that incorporates the effect of color decoherence, and
reads (for a homogeneous medium) 
\begin{align}
\label{eq:decoh-parameter}
\Delta_\text{med}(t) &= 1-\frac{1}{N_c^2-1} \tr_c \left[ \langle \med|  U_\bkg(n_1) U_\bkg^\dag(n_2) | \med\rangle \right]\,,\\
&\approx 1- \exp \left[ - \frac{1}{12} \hat q \, \theta^2_{12}\,  t^3 \right]  \,.
\end{align}
To obtain the last line, the harmonic-oscillator approximation, which resums multiple soft scatterings, has been employed. Systematic approaches that go beyond this approximation have since been developed~\cite{Kuzmin:2025fyu}, notably the Improved Opacity Expansion (IOE) and the Improved Harmonic Oscillator (IHO) framework~\cite{Mehtar-Tani:2019tvy,Mehtar-Tani:2019ygg}. These methods incorporate hard scatterings on top of a background of multiple soft interactions. Recent works have also investigated the role of the finite formation time in the shape of the decoherence parameter $\Delta_{\rm med}$~\cite{Abreu:2024wka}, the impact of plasma anisotropies~\cite{Barata:2024bqp} or the interplay between $\Delta_{\rm med}$ and the transverse momentum broadening of the antenna~\cite{Arleo:2026urv}. We refer the reader to the now extensive literature for a comprehensive discussion of these developments.

 \section{Quantum corrections to color decoherence}
 \label{sec:thetac_evol}
The coherence angle plays a central role in the physics of medium-induced energy loss. It characterizes the transition between coherent and incoherent energy-loss regimes. At leading order in perturbation theory, a two-pronged system in an arbitrary color representation \(R\), such as a quark--antiquark antenna, loses energy as a single color charge with Casimir \(C_R\) when the opening angle satisfies \(\theta_{12}<\theta_c\). In this regime, the medium is unable to resolve the internal structure of the antenna. For an arbitrarily narrow color-singlet antenna, the energy loss vanishes,
\beq
\bar S_{12}^{\rm med}(\epsilon)\simeq \delta(\epsilon)\,,
\eeq
or equivalently,
\beq
\bar S_{12}^{\rm med}(\nu)=1\,,
\eeq
in Laplace space. This phenomenon is known as \emph{color transparency}.

Conversely, when \(\theta_{12}>\theta_c\), the medium resolves the antenna substructure and the two prongs lose energy independently. The total energy loss is then governed by the sum of the individual color charges, leading to
\beq
\bar S_{12}^{\rm med}(\epsilon)\simeq
\int \rmd\epsilon_1
\int \rmd\epsilon_2\,
\bar S_1^{\rm med}(\epsilon_1)\bar S_2^{\rm med}(\epsilon_2)\,
\delta(\epsilon-\epsilon_1-\epsilon_2)\,,
\eeq
or, in Laplace space,
\beq
\bar S_{12}^{\rm med}(\nu)=\bar S_1^{\rm med}(\nu)\bar S_2^{\rm med}(\nu)\,.
\eeq
The coherence angle \(\theta_c\) therefore defines the angular scale at which the medium transitions from seeing the antenna as a single color charge to resolving its individual constituents. In the multiple-soft scattering (harmonic oscillator, HO) approximation for a sufficiently large medium, it is given parametrically by
\beq \label{eq:theta_c0}
\theta_c(0) \sim \frac{1}{\sqrt{\hat q L^3}}\,.
\eeq

\subsection{Asymptotics of the color coherence angle}

Equation~\eqref{eq:theta_c0} is a leading-order result without vacuum quantum evolution. One of the central findings of this work is that the resummation of large \(\ln N\) contributions associated with multiple soft-collinear emissions inside the jet further reduces the decoherence angle. In other words, radiative corrections enhance the resolving power of the medium, causing color coherence to be lost at smaller angular separations. Indeed, although the decoherence angle was originally introduced in the context of medium-induced parton energy loss, closely related dynamics also arise in the vacuum BMS evolution. In particular, the emergence of a critical angular scale, governed by the evolution variable
\begin{align}
\Delta \sim \bar\alpha_s\ln \frac{E}{E_{\rm loss}} \sim \bar\alpha_s\ln N\,,
\end{align}
was first identified in Ref.~\cite{Banfi:2002hw}. In this context, the critical angle $\theta_c(\Delta)$ controls the effective angular size of the active jet region which contributes to the jet energy loss, i.e.~the physical opening angle of the jet. The physical picture is simple: the wider the antenna configuration inside the jet, the larger the phase space available for radiation to escape the jet region. Consequently, as $E_{\rm loss}$ decreases (and $\Delta$ increases), it becomes increasingly favorable to suppress such wide-angle configurations in order to prevent them from radiating outside the jet. 
Above $\theta_c(\Delta)$ up to the jet radius $R$, real gluon emissions are suppressed, forming an empty buffer region with virtual emissions only~\cite{Dasgupta:2002bw}. As $\ln(E/E_{\rm loss})$ increases, $\theta_c(\Delta)$ decreases, so that the jet becomes effectively narrower and narrower, irrespective of its initial geometrical size set by $R$. This universality of the jet shape at large $\ln N$ is analogous to the universality of the dipole $S$-matrix or the unintegrated dipole gluon distribution at small $x$. In both cases, an emergent scale governs the asymptotic behavior independently of the initial condition: the saturation scale $Q_s(x)$ controls the $k_T$ dependence of the unintegrated gluon distribution, while its counterpart in the BMS context is $\theta_c(\Delta)$.

To be slightly more quantitative, in vacuum, the critical angle decreases exponentially with the evolution variable~\cite{Neill:2016stq},
\beq
\theta_c(\Delta) \sim R\, e^{-c\Delta}\,,
\eeq
where \(R\) is the jet radius and \(c\) is a numerical constant determined by the asymptotic solution of the BMS equation. As we have just seen, medium interactions provide an additional mechanism for resolving the jet substructure. The role of the jet radius is played by the leading-order decoherence angle,
\begin{align}
R \;\rightarrow\; \theta_c(0)\,,
\end{align}
with \(\theta_c(0)\) given by eq.\,\eqref{eq:theta_c0}. The resummed decoherence angle therefore inherits the same exponential suppression,
\beq
\theta_c(\Delta) \sim \theta_c(0)\, e^{-c\Delta}\,,
\eeq
demonstrating that successive soft-collinear emissions progressively reduce the angular scale over which color coherence is preserved.
This parametric discussion is intended to emphasize the unification of the theoretical frameworks describing non-global logarithms and jet quenching. Despite their different physical settings, both phenomena exhibit the emergence of a dynamically generated critical angle associated with soft radiation, revealing a common underlying structure. This observation constitutes the central physics result of this work.

The parallel between the jet-quenching problem considered here and saturation physics at small $x$ can be fruitfully exploited to derive the universal large-$\Delta$ behavior of the decoherence angle. By universal, we mean --- just as in the saturation case --- that the asymptotic behavior is independent of the initial condition. A particularly attractive feature of the BMS equation is that it can be readily generalized to include running-coupling effects through an appropriate redefinition of the evolution variable. At one loop, the evolution variable $\Delta$ is defined as~\cite{Banfi:2002hw,Becher:2016mmh} 
\begin{align}
    \Delta = \frac{N_c}{\pi}\int_{R/\nu}^{p_TR}\frac{\rmd\mu}{\mu}\alpha_s(\mu)=-\frac{N_c}{\pi\beta_0}\ln(1-\alpha_s(p_TR)\beta_0\ln(\nu p_T))\label{eq:Delta-rc-def}
\end{align}
with $\beta_0=(11 N_c-2n_f)/(6\pi)$ for $n_f$ active quark flavors. Expanding $\Delta$ for $\alpha_s(p_TR)\ll 1$, one recovers the parametric estimate $\Delta \sim\bar\alpha_s\ln(\nu p_T)$  which is equal to $\bar\alpha_s\ln(N)$ for $\nu=N/p_T$. With this evolution variable, the BMS equation eq.\,\eqref{eq:bms-12} for $\bar S_{12}$ in the limit $R\to\infty$ can be mapped onto the BK equation 
\begin{align}
    \frac{\rmd \bar S_{12}}{\rmd \Delta}&=\int\frac{\rmd^{2}\btheta_{3}}{2\pi}\,
\frac{\btheta_{12}^2}{\btheta_{13}^2 \btheta_{32}^2}
\left[
\bar S_{13} \bar S_{32} -\bar S_{12}
\right] \, .\label{eq:BK-largeR}
\end{align}
The limit $R\to\infty$ is justified a posteriori by noting that the emergent angular scale $\theta_c(\Delta)$, conventionally defined by the implicit equation
\begin{align}
    \bar S_{12}(\theta_c(\Delta),\Delta)=\frac{1}{2}\,,
\end{align}
is much smaller than $R$ at large $\Delta$ (both in the vacuum and medium cases), such that the integral is largely controlled by angles $\ll R$.

Since the universal small-$x$ behavior of the saturation scale $Q_s(x)$, as determined by the fixed-coupling BK equation, is already known~\cite{Mueller:2002zm,Munier:2003vc,Munier:2003sj}, the asymptotic large-$\Delta$ behavior of $\theta_c(\Delta)$ can be readily inferred:
\begin{align}
    \ln(\theta_c(\Delta))=-c \Delta+b\ln(\Delta)+\textrm{const}+\mathcal{O}\left(\frac{1}{\sqrt{\Delta}}\right)\,,\label{eq:thetac-asymptotic}
\end{align}
with $c$ and $b$ two numerical constants. $c$ is the velocity of the front associated with the $\theta_c$ transition and is given by $c=\chi(\gamma_c)/(2\gamma_c)$ where $\gamma_c$ is the unique solution to the implicit equation $\chi(\gamma_c)=\gamma_c\chi'(\gamma_c)$ with $\chi(\gamma)=2\psi(1)-\psi(\gamma)-\psi(1-\gamma)$ the LO BFKL eigenvalue~\cite{Kuraev:1977fs,Balitsky:1978ic}. (We note $\psi(z)$ the digamma function.) The second subleading coefficient $b$ is given by $b=3/(4\gamma_c)$. Numerically, one has $c\simeq 2.44$ and $b\simeq 1.20$. The constant term in eq.\,\eqref{eq:thetac-asymptotic} is non-universal and set by the initial condition: $\textrm{const}\sim (\hat q L^3)^{-1/2}$ in the medium case and $\textrm{const}\sim R$ in the vacuum. Note that the $\mathcal{O}(1/\sqrt{\Delta})$ correction is also universal, and given by $3\gamma_c^{-2}\sqrt{\pi/(2\chi''(\gamma_c))}\times\Delta^{-1/2}\simeq 1.37\Delta^{-1/2}$.

Beyond the asymptotic behavior of the coherence angle including quantum corrections, the correspondence between (de)coherence in jet quenching and small-$x$ saturation also provides analytic control over the BMS solution in the vicinity of the coherence transition at large $\Delta$. Using the results of~\cite{Munier:2003sj,Munier:2003vc}, one finds
\begin{align}
\bar S_{12}(x=\theta_{12}/\theta_c(\Delta),\Delta)&\simeq \sqrt{\frac{2}{\chi''(\gamma_c)}}x^{-2\gamma_c}\ln x \times \exp\left(-\frac{2}{\chi''(\gamma_c)\Delta}\ln^2(x)\right)\,.\label{eq:front-transition-S12}
\end{align}
This expression is formally valid in the asymptotic large-$\Delta$ regime. All medium effects are encoded in the initial condition through $\theta_c(\Delta=0)$. Although phenomenologically relevant values of $\tau$ do not lie in the asymptotic regime, eq.\,\eqref{eq:front-transition-S12} provides a fully analytic description of the interplay between medium-induced energy loss and the vacuum parton shower, and may serve as a useful starting point for phenomenological model studies.

\subsection{In-medium one-prong soft-collinear function}

We wish now to analytically study the one-prong soft-collinear function $\bar S_1(\Delta,\boldsymbol{\theta}_1)$ after BMS evolution and using an initial condition set by the medium. For simplicity, we focus on $\bar S_1(\Delta,\boldsymbol{\theta}_1=\boldsymbol{0})$ evaluated at vanishing angle such that the initial parton sourcing the jet propagates along the $\theta=0$ axis. Using the very weak dependence of $\bar S_1(\Delta,\boldsymbol{\theta}_1)$ with $\boldsymbol{\theta}_1$, we can approximate $\bar S_1(\Delta,\boldsymbol{\theta}_1)\approx \bar S_1(\tau,\boldsymbol{0}_\perp)$ in the right hand side of the BMS equation for $\bar S_1$. After these simplifications, we find the equation\footnote{As compared to eq.\,\eqref{eq:bms-full-large-Nc}, we include the $R$-dependent logarithm in the evolution equation for $S_1$, rather than factoring it out into $S_{\rm tot}$, in order to recover the standard Sudakov factor in the double logarithmic approximation, cf eq.\,\eqref{eq:S1vac-dla}.}:
\begin{align}
    \frac{\rmd \bar S_1(\Delta)}{\rmd\Delta} &\simeq\ln(R) \bar S_1(\Delta)+ \bar S_1(\Delta )\int_{|\boldsymbol{\theta}'|\le 1}\frac{\rmd^2\boldsymbol{\theta}'}{2\pi}\frac{1}{\boldsymbol{\theta}'^2}\left[ \bar S_{12}(\Delta,|\boldsymbol{\theta}'|)-1\right]\,,\\
    &=\ln(R) \bar S_1(\Delta)- \bar S_1(\Delta)\int_0^R \frac{\rmd \theta'}{\theta'}(1- \bar S_{12}(\Delta,\theta))\,.\label{eq:nl-s1-approx}
\end{align}
We have renamed $\bar S_1(\Delta,\boldsymbol{\theta}_1=\boldsymbol{0})\to \bar S_1(\Delta)$ for the sake of simplicity. This expression can be further estimated using a stepwise approximation for the $\theta$ dependence of $\bar S_{12}(\Delta,\theta)$:
\begin{align}
     \bar S_{12}(\Delta,\theta)= \bar S_1^2(\Delta)\,,
\end{align}
if $\theta>\theta_c(\Delta)$ and $ \bar S_{12}(\Delta,\theta)=1$ otherwise. 

This piecewise approximation for $S_{12}$ enables one to close the differential equation satisfied by $ S_1$. We find indeed
\begin{align}
    \frac{\rmd\bar  S_1(\Delta)}{\rmd\Delta} 
    &=\ln(R) \bar S_1(\Delta)-\ln(R/\theta_c(\Delta)) \bar S_1(\Delta)(1- \bar S_1^2(\Delta))\,,\\
    &=\ln(\theta_c(\Delta)) \bar S_1(\Delta)+\ln(R/\theta_c(\Delta)) \bar S_1^3(\Delta)\,.\label{eq:S1-med-approx}
\end{align}
In the absence of medium effects, using $\theta_c(\Delta)\simeq \theta_c(0)=R$, the previous manipulations would simply lead to the linear differential equation
\begin{align}
     \frac{\rmd \bar S_1(\tau)}{\rmd\Delta} 
    &=\ln(R) \bar S_1(\Delta)\,,
\end{align}
whose solution with the initial condition $ S_1(0)=1$ is the standard Sudakov suppression factor in the double logarithmic approximation
\begin{align}
     \bar S_1(\Delta)&=\exp(-\Delta\ln(1/R))\simeq \exp(-\bar\alpha_s\ln(N)\ln(1/R))\,,\label{eq:S1vac-dla}
\end{align}
In the presence of the medium, let us first consider the linearized version of eq.\,\eqref{eq:S1-med-approx} and neglecting the $\Delta$ dependence of $\theta_c$. The solution is then 
\begin{align}
     \bar S_1(\Delta)&\simeq \exp(-\bar\alpha_s\ln(N)\ln(1/\theta_c))\,,
\end{align}
which agrees with the result obtained from the nonlinear DGLAP evolution in the double-logarithmic approximation~\cite{Mehtar-Tani:2024mvl}. It shows again, as compared to eq.\,\eqref{eq:S1vac-dla}, that the coherence effects from the medium leads to the replacement $R\to\theta_c$ in the physical opening angle of the jet. It is then interesting to investigate the effect of the non-linear term. The general solution to eq.\,\eqref{eq:S1-med-approx} is
\begin{align}
     \bar S_1(\Delta)&=\frac{ \bar S_1(0)}{\sqrt{1-2f(\Delta) \bar S_1^2(0)}}\exp\left(\int_0^\Delta \rmd \Delta' \ln[\theta_c(\Delta')]\right)\,,
\end{align}
with
\begin{align}
    f(\Delta)&=\int_0^\Delta \rmd\Delta' \ln[R/\theta_c(\Delta')]\exp\left(2\int_0^{\Delta'} \rmd \Delta'' \ln[\theta_c(\Delta'')]\right)\,.
\end{align}
Neglecting first the $\Delta$ dependence of $\theta_c$, one gets
\begin{align}
    \bar S_1(\Delta)&=\frac{ \bar S_1(0)e^{-\ln(1/\theta_c)\Delta}}{\sqrt{1-\frac{\ln(R/\theta_c) \bar S_1^2(0)}{\ln(1/\theta_c)}\left(1-e^{-2\ln(1/\theta_c)\Delta}\right)}}\,.
\end{align}
If one uses the expression for $\theta_c(\Delta)\simeq\theta_c(0)e^{-c\Delta}$ obtained in the previous subsection, then
\begin{align}
     \bar S_1(\Delta)&=\frac{\bar  S_1(0)e^{-\ln(1/\theta_c(0))\Delta-c\Delta^2/2}}{\sqrt{1-2f(\Delta) \bar S_1^2(0)}}\,,
\end{align}
with 
\begin{align}
    f(\Delta)&=\frac{1-e^{2\ln(\theta_c(0))\Delta-c\Delta^2}}{2}\nonumber\\
    &+\frac{e^{\ln^2(\theta_c(0))/c}\ln(R)\sqrt{\pi}}{2\sqrt{c}}\left[\textrm{erf}\left(\frac{\ln(\theta_c(0))}{\sqrt{c}}\right)+\textrm{erf}\left(\frac{c\Delta-\ln(\theta_c(0))}{\sqrt{c}}\right)\right]\,.
\end{align}
Here $\textrm{erf}(z)$ is the error function.

\subsection{Implications for inclusive jet suppression}

\begin{figure}
    \centering
    \includegraphics[width=0.5\linewidth]{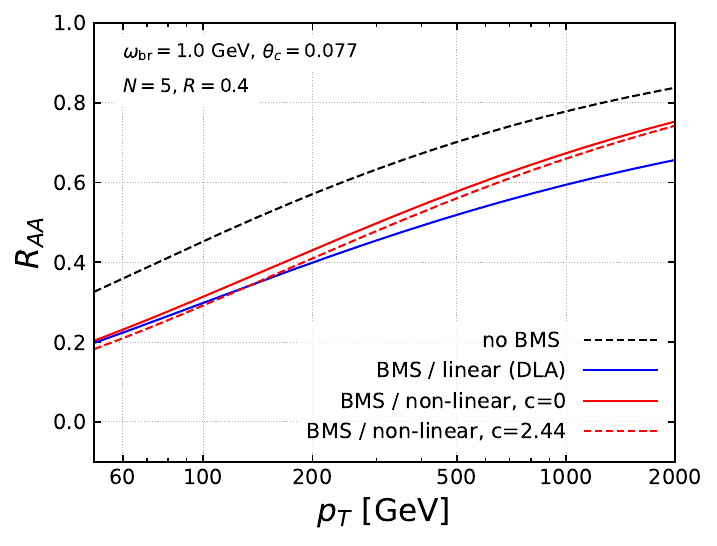}
    \caption{Nuclear modification for jets as obtained from the approximate solution to the BMS evolution of the medium initial condition.}
    \label{fig:RAA-from-BMS}
\end{figure}

The above expression can be used to obtain an estimate of the nuclear modification factor. Without non-linear effects, one simply gets
\begin{align}
    R_{AA}(p_T)&= \bar S_1(0)\times \exp\left(-\ln(R/\theta_c)\Delta\right)\,,\label{eq:RAA-DLA-BMS}
\end{align}
which is the standard DLA results that can also be obtained using collinear evolution instead of BMS (see e.g.~\cite{Mehtar-Tani:2024mvl}). When the non-linear term is included, on gets
\begin{align}
    R_{AA}(p_T)&=\frac{ \bar S_1(0)e^{-\ln(R/\theta_c(0))\Delta-c\Delta^2/2}}{\sqrt{1-2f(\Delta) \bar S_1^2(0)}}\,.\label{eq:RAA-BMS}
\end{align}
The initial condition incorporates medium-induced energy loss and for the plot in figure~\ref{fig:RAA-from-BMS}, we use the BDMPS-Z result $ \bar S_1(0)=\bar S_1^{\rm med}=\exp(-2\sqrt{\pi \omega_{\rm br}\nu})$ with $\omega_{\rm br}\sim\bar\alpha_s^2\qhat L^2$. This corresponds to the exponentiation of the result obtained in section~\ref{sec:med-boundary}, in particular eq.\,\eqref{eq:S1-BDMPS} ; the exponentiation describing multiple independent medium-induced gluon emissions. The nuclear modification factor obtained from this initial condition $ S_1(0)$ (with $\nu=N/p_T$) is shown on the black dashed curved in figure~\ref{fig:RAA-from-BMS} for $\omega_{\rm br}=1$ GeV. The blue curve corresponds to eq.\,\eqref{eq:RAA-DLA-BMS}, which is the Sudakov suppression including coherence effects in the linear regime of the BMS equation. We use $\theta_c(0)=0.077$. Note that the Sudakov factor has a (mild) $p_T$ dependence coming from $\Delta$ given by eq.\,\eqref{eq:Delta-rc-def} when running coupling effects are included. Finally, the red curves are the solutions given by eq.\,\eqref{eq:RAA-BMS} with (dashed) or without (plain) the $\Delta$-dependence of $\theta_c$ through the relation $\theta_c=\theta_c(0)e^{-c\Delta}$. The effect of the $\Delta$ dependence is very small for the phenomenological values of $\Delta\sim 0.2$ probed by the plot which are far from the asymptotic regimes $\Delta\gg 1$. Yet, the effect of the non-linear term is substantial at large $p_T$ where it tames the suppression of the jet cross-section. This can easily be understood from eq.\,\eqref{eq:RAA-BMS} which shows that the non-linear contribution is important when we do not have $ S_1(0)\ll1$, meaning $p_T\gg N\omega_{\rm br}$. The correspondence between jet quenching and saturation physics thus offers an opportunity to measure non-linear QCD effects non only in small $x$ processes but also in high-$p_T$ jet suppression in heavy-ion collisions.

\section{Wilson-Line Correlators in the Background Field Formalism}
\label{sec:Wilson-correlators-eft}
This section is more formal in nature and serves as a complement to the main body of this work. In particular, we demonstrate how effective field theory techniques can be employed to reproduce and systematically derive known results for medium-induced radiation. It will also establish an explicit connection between the Color Glass Condensate (CGC)~\cite{Gelis:2010nm} formalism and the theory of jet quenching.

\subsection{Mode separation and power counting}

The gauge field entering the Wilson lines can be naturally separated into fast and slow degrees of freedom according to their light-cone momentum component $k^+$. To implement this separation, we introduce an intermediate factorization scale $\Lambda^+$ satisfying
\beq
\frac{p_T}{N} \gg \Lambda^+ \gg T \, ,
\eeq
and decompose the gauge field as
\beq\label{eq:field-fact}
A^\mu(x)
=
a^\mu(x;k^+>\Lambda^+)
+
A^\mu_{\rm bkg}(x;k^+<\Lambda^+) \, .
\eeq
The field $a^\mu$ describes the fast quantum fluctuations associated with collinear-soft radiation, while $A^\mu_{\rm bkg}$ represents a slowly varying background field generated by the medium. Physically, the latter encodes the collective long-wavelength dynamics of the plasma and mediates soft momentum exchanges between the jet and the medium constituents.

The typical scaling of the collinear-soft modes is
\beq
k_{\rm cs}
\sim
p_T\beta\left(1,\delta^2,\delta\right) .
\eeq
The medium background modes, which we identify with Glauber exchanges, differ from the collinear-soft modes primarily through an additional suppression of the $+$ momentum component. Parametrically, we take
\beq
k_{\rm bkg}
\sim
p_T\beta\left(\delta,\delta^2,\delta\right) .
\eeq
Thus, the Glauber and collinear-soft modes share the same transverse and $-$ momentum scaling and differ only in their longitudinal $+$ momentum component. Because of this overlap in scaling, the Glauber modes cannot be systematically integrated out or factorized from the collinear-soft sector.

This scaling can be understood from a simple kinematic argument. Consider a $2\to2$ scattering process between a collinear-soft parton with momentum
\beq
k_{\rm cs}\sim (k_{\rm cs}^+,0,0_\perp)
\eeq
and a thermal plasma constituent with momentum
\beq
k_{\rm pl}\sim (0,T,0_\perp) .
\eeq
Let $k_{\rm bkg}$ denote the exchanged momentum. Imposing the on-shell conditions for the outgoing particles gives
\beq
(k_{\rm cs}+k_{\rm bkg})^2
\simeq
2k_{\rm cs}^+k_{\rm bkg}^-
-k_{\rm bkg\perp}^2
\simeq 0 ,
\eeq
and
\beq
(k_{\rm pl}-k_{\rm bkg})^2
\simeq
-2k_{\rm pl}^-k_{\rm bkg}^+
-k_{\rm bkg\perp}^2
\simeq 0 .
\eeq
These relations imply
\beq
k_{\rm bkg}^-
\sim
\frac{k_{\rm bkg\perp}^2}{k_{\rm cs}^+}
\sim
p_T\beta\delta^2 ,
\eeq
and
\beq
k_{\rm bkg}^+
\sim
\frac{k_{\rm bkg\perp}^2}{k_{\rm pl}^-}
\sim
\frac{p_T}{T}\beta^2\delta^2 .
\eeq
Using the thermal scaling
\beq
T\sim p_T\beta\delta ,
\eeq
one finds
\beq
k_{\rm bkg}^+
\sim
p_T\beta\delta ,
\eeq
which reproduces the Glauber scaling quoted above.

We now turn to the structure of the Wilson line,
\beq\label{eq:WL-G-SC}
U_n
=
P\exp\left[
ig\int_0^1\rmd s\,
n\cdot\left(
A_\bkg(sn)
+
a(sn)
\right)
\right] .
\eeq
The contribution from the collinear-soft field can be decomposed as
\beq
n\cdot a(sn)
=
n^+a^-(sn)
+
n^-a^+(sn)
-
{\bs n}_\perp\cdot{\bs a}_\perp(sn) .
\eeq
Using the scaling relation $a^\mu\sim k_{\rm cs}^\mu$, we obtain
\beq
a^+\sim 1,
\qquad
a^-\sim \delta^2,
\qquad
a_\perp\sim \delta .
\eeq
Together with
\beq
n^+\sim 1,
\qquad
n^-\sim \delta^2,
\qquad
n_\perp\sim \delta ,
\eeq
this implies
\beq
n^+a^-
\sim
n^-a^+
\sim
{\bs n}_\perp\cdot{\bs a}_\perp
\sim
\delta^2 .
\eeq
Hence all components contribute at the same parametric order. This is simply a consequence of the approximate Lorentz symmetry underlying the collinear-soft scaling.

The situation is qualitatively different for the Glauber background field. Since the medium defines a preferred rest frame, Lorentz symmetry is explicitly broken, and the different components of the background field no longer contribute democratically. In particular, the dominant contribution arises from the $A^-_\bkg$ component.

To see this, consider first the limit of a static medium composed of nonrelativistic color sources. The corresponding classical field configuration takes the form
\beq
A^\mu_\bkg \to A_\bkg^0(\vec{x}),
\qquad
\vec{A}_\bkg=0 .
\eeq
Expressed in light-cone coordinates, this implies
\beq
A^-_\bkg
\sim
A^+_\bkg
\sim
A^0_\bkg ,
\qquad
A_{\bkg\perp}=0 .
\eeq
The contraction with the collinear direction therefore gives
\beq
n^+A^-_\bkg
\gg
n^-A^+_\bkg
\sim
\delta^2 A^-_\bkg .
\eeq
Thus the $A^-_\bkg$ component dominates at leading power.

The same conclusion holds even in a dynamical plasma where all field components are parametrically comparable,
\beq
A^+_\bkg
\sim
A^-_\bkg
\sim
A_{\bkg\perp} .
\eeq
In this case one finds the hierarchy
\beq
n^+A^-_\bkg
\gg
\delta A_{\bkg\perp}
\gg
\delta^2 A^+_\bkg .
\eeq
Therefore, independent of the microscopic details of the medium, the leading interaction is controlled by the $A^-_\bkg$ component.

A similar simplification occurs in the coordinate dependence of the Glauber field. For collinear-soft modes, all components of the coordinate scale uniformly and must be retained. For Glauber modes, however, the Fourier phase behaves as
\beq
x\cdot k_\bkg
=
s\,n\cdot k_\bkg
=
s\left(
n^+k^-_\bkg
+
n^-k^+_\bkg
-
{\bs n}_\perp\cdot{\bs k}_{\bkg\perp}
\right)
\sim
s\left(
\delta^2
+
\delta^3
+
\delta^2
\right) .
\eeq
The term involving $k^+_\bkg$ is therefore power suppressed and can be neglected at leading order. As a result,
\beq
A^-_\bkg(sn)
\simeq
A^-_\bkg(sn)\Big|_{n^-=0}
=
A^-_\bkg(x^+,x_\perp)\Big|_{x=sn} .
\eeq

The Wilson line thus reduces to
\beq\label{eq:WL-G-SC-3}
U_n
\simeq
P\exp\left[
ig\int_0^1\rmd s
\left(
A^-_\bkg(sn)\Big|_{n^-=0}
+
n\cdot a(sn)
\right)
\right]
\left(1+\mathcal{O}(\delta)\right) .
\eeq

This eikonal approximation is the basis of high energy factorization underling small-$x$ physics.

\subsection{The LO and NLO order in the perturbation $a^\mu$}

As a proof of concept, let us compute the leading radiative correction to the soft function $S_m$ for a single collinear mode. Namely, we shall consider (recall that $U_0=1$ in the $A^+=0$ light-cone gauge)
\beq \label{eq:soft-function-1}
&&S_1 (\epsilon,R) \equiv \frac{1}{N_c} {\rm tr}_c \sum_{X}   \, \delta(\epsilon - \bar n \cdot p_{\rm out} ) \,  \langle {\rm med}|  U_1^\dag(n) | X \rangle  \langle X| U_{1}(n)| {\rm med} \rangle  \,
 \eeq
 \subsection{Soft-collinear function at leading order}
 To zeroth order in the soft-collinear field and all order in the Glauber we have 
 \beq 
&&S^{(0)}_1 (\epsilon,R) \equiv \sum_{X}   \, \delta(\epsilon - \bar n \cdot p_{\rm out} ) \,  \langle {\rm med}|  U_{\bkg, 1}^\dag(n) | X \rangle  \langle X| U_{\bkg, 1}(n)| {\rm med} \rangle  \,. 
 \eeq
However,  we can neglect the longitudinal momentum transfer of the Glauber which is of order $\delta \beta \ll \beta $ (this approximation can be relaxed easily leading to collisional energy loss) . $\beta$ being the radiative energy loss scale. In this situation we can write 
\begin{align}
 S^{(0)}_1 (\epsilon,R)  &=  \frac{1}{N_c} {\rm tr}_c \sum_{X}   \, \delta(\epsilon  ) \,  \langle {\rm med}|  U_{\rm bkg, 1}^\dag(n) | X \rangle  \langle X| U_{\rm bkg, 1}(n)| {\rm med} \rangle (1+O(\delta \beta))\,, \\
  &= \langle {\rm med}|  U_{\rm bkg, 1}^\dag(n) U_{\rm bkg, 1}(n)| {\rm med} \rangle(1+O(\delta \beta))\,, \\
    &\approx  1 \,,
\end{align}
where we have used the completeness relation   $\sum_{X}  | X \rangle  \langle X|  =1$. 

Before we delve into higher order corrections we anticipate on the fact that we are interested in the medium-induced corrections that are enhanced by the medium length and the early time vacuum radiation that both occur at time scales of order $x^+\sim (\beta \delta^2)^{-1} \ll L$.  

\subsection{Real radiation in the background field to all orders }

At the next order, we expand the Wilson line to linear  order in $a^-$ (the  quadratic order will be necessary for the virtual contribution) as depicted in Figure~\ref{fig:single-med-emission}: . That is, 
\begin{align}
     U_1(+\infty, 0) &\equiv \cP \exp\left[ ig \int_0^{+\infty} \rmd x^+ \left( A^- _{\rm bkg}(x^+,0^-,0_\perp) +a^-(x^+,0^-,0_\perp)\right)\right]\,,\\
     &\simeq i g \int_0^\infty \rmd z^+ U_{\rm bkg, 0_\perp} (+\infty, z^+)   a^-(z^+,0^-,0_\perp)U_{\rm bkg, 0_\perp} (z^+,0) \,.
\end{align}
The real contribution at leading order yields 
\begin{align}
S^{(1)}_1 (\epsilon,R)  &\approx   \frac{g^2 }{N_c} {\rm tr}_c \sum_{X}   \,  \,  \langle \med|  \int_0^{+\infty} \rmd z_2^+  U^\dag _{\rm bkg} (z_2^+,0)  a^-(z_2^+) U^\dag_{\rm bkg} (+\infty, z_2^+)   | X \rangle  \delta(\epsilon -\omega  ) \nonumber\\
 &\times  \langle X|   \int_0^{+\infty}\rmd z_1^+ U_{\rm bkg} (+\infty, z^+)   a^-(z_1^+)U_\bkg (z^+,0) )| \med \rangle \Big|_{x_\perp=0_\perp}  \,,\\
 &=    \frac{g^2 }{N_c}   \sum_{X}  \int_0^{+\infty} \rmd z_2^+  \int_0^{+\infty} \rmd z_1^+   \delta(\epsilon -\hat\omega  ) \nonumber\\
 &\times\langle \med| a^{b,-}(z_1^+) \delta(\epsilon -\hat\omega  )  |X\rangle \langle X| {\rm tr}_c  \left(t^a[z_1^+, z_2^+] t^b[z_2^+, z_1^+]\right) a^{a,-}(z_1^+)| \med \rangle  \,.
\end{align}
This has the form of an amplitude squared. Assuming that the Glauber gluons do not contribute to the final state, the corresponding Wilson lines cancel out outside the time interval $[z^+_1,z^+_2]$. Here we have used the symbolic notation $\hat \omega$ that denotes an operator that measures the energy radiated out of the cone. 

Using the following algebraic identity we may further simplify by combining the two fundamental Wilson lines into an adjoint one:
\beq
 [z_2^+, z_1^+] t^a[z_1^+, z_2^+] = t^b \cU^{ba} (z_2^+, z_1^+) \,.
\eeq
As a result we have 
\beq
 {\rm tr}_c  \left(t^a[z_1^+, z_2^+] t^b[z_2^+, z_1^+]\right) =\frac{1}{2} \cU^{ba} (z_2^+, z_1^+) \,.
\eeq
To compute the correlator of the collinear-soft gauge field, we need their propagator $G$ in the presence of the medium (throughout we use the light-cone gauge $A^+=0$)
\beq
&&a^{b,-}(z_1^+) \delta(\epsilon - k^+_{\rm out}  )  |k_{\rm out}\rangle \langle k_{\rm out}| a^{a,-}(z_2^+) 
\nn &&\to \int \rmd^{d} k (2\pi) \delta(k^2)\theta(k^+) \delta(\epsilon - k^+ )\Theta_{\rm alg} \nn
&&  \times k^2\int \rmd^d y \rme^{-i k\cdot y }  G^{\dag, cb, -}_{\mu}( y, z_2^+)   k^2  \int \rmd^d x \rme^{i k\cdot x } G^{ca,\mu -}( x, z_1^+)\,.
\eeq
Applying the following reduction formula for dressed external propagators, we obtain
\begin{align}
 \lim_{k^2\to 0 } k^2 \int \rmd^d k_0 & \ G_{\rm scal}(k,k_0) J(k_0)\\
 &= \rme^{i \frac{\bk^2}{2 k^+} L_{\infty}} \int \rmd^{d-2} \bk_0  \int \rmd z^+_0 (\bk| \cG(L_{\infty}, z^+_0) |\bk_0)  J(\bk_0,k^+, z^+_0)\,,
 \end{align}
 for any test source function $J(k_0)$, and therefore
 \begin{align}
S^{(1)}_1 (\epsilon,R)  &\approx    \frac{g^2}{2 N_c}  \int \rmd^{d} k (2\pi) \delta(k^2)\theta(k^+) \delta(\epsilon - k^+ )\Theta_{\rm alg} \nn
 &  \frac{\bk_1\cdot \bk_2}{(k^+)^2}  \cU^{ba} (z_1^+, z_2^+)  (\bk_2| \cG^{bc}(z^+_2, L_{\infty}) |\bk)    (\bk| \cG^{ca}(L_{\infty}, z^+_1) |\bk_1)  \nn 
 &  \to \frac{\bk_1\cdot \bk_2}{(k^+)^2}  {\rm tr}_c (\cU^\dag (z_1^+, z_2^+)  (\bk_2| \cG^\dag(z^+_2, L_{\infty}) |\bk)    (\bk| \cG(L_{\infty}, z^+_1) |\bk_1) \,.
 \end{align}
 At this stage, one may invoke the fact that Glauber gluon exchanges are quasi-instantaneous  compared to the time scale of the collinear-soft radiation. This allows us to factorize the above probability into a radiation part between $z^+_1$ and $z^+_2$ and a propagation part from $z^+_2$ to $+\infty$ for $z^+_2>z_1^+$ and similarly for the opposite ordering which can be accounted for by taking 2 times the real part of the contribution $z^+_2>z_1^+$,
 \begin{align}\label{eq:eloss-prob-1}
 S^{(1)}_1 (\epsilon,R)  &\approx    \frac{g^2}{2 N_c}  \int \rmd^{d} k (2\pi) \delta(k^2)\theta(k^+) \delta(\epsilon - k^+ )\Theta_{\rm alg} \nn
 &\times 2 {\rm Re} \int _0^{+\infty}  \rmd {z^+_2}   \int_0^{z^+_2} \rmd z^+_1 \cP(+\infty, z_2^+; \bk,\bq)  \cK(z_2^+, z_1^+;\bq,\bk_1)  \,,
 \end{align}
 where we define, upon using translational invariance of the medium in the transverse plane,
 \begin{align}
 (2\pi)^2 \delta(\bk_2-\bq) \cP(z_2^+, z_1^+;\bk,\bq) = \frac{1}{N_c^2-1}  {\rm tr}_c  \langle    (\bk| \cG^\dag(+\infty, z^+_2) |\bk_2)   (\bk| \cG(+\infty, z^+_2) |\bq)   \rangle   \,,
 \end{align}
 and 
  \begin{align}
\cK(z_2^+, z_1^+; \bq,\bk_1)=   \frac{(\bq\cdot \bk_1)}{(k^+)^2} \, {\rm tr}_c  \langle   \,  \cU^\dag(z_2^+, z_1^+)  (\bq| \cG(z_2^+, z_1^+) |\bk_1)    \rangle \,,  
 \end{align}
 where the propagator $ \cG$ can be viewed as the time evolution operator of quantum particle in the presence of a potential in 2+1 dimensions and therefore obeys the Schr\"odinger equation 
 \beq\label{eq:cG-prop}
\left[i\frac{\del}{\del z^+} + \frac{\del^2_\perp}{2 k^+}+gA_\bkg(z^+,\bz) \right] (\bz|\cG(z^+,z^+_0)|\bz_0)=i\delta(z^+-z^+_0)\delta(\bz-\bz_0)\,.
\eeq
In integral form, it reads 
 \begin{align}\label{eq:cG-prop-2}
(\bz|\cG(z^+,z^+_0)|\bz_0)&=(\bz|\cG_0(z^+,z^+_0)|\bz_0)\nonumber\\
&+ig (\bz_1|\cG_0(z^+,z^+_1)|\bz_1)A_\bkg(z_1^+,\bz_1)(\bz_1|\cG(z_1^+,z^+_0)|\bz_0) \,.
\end{align}
At one loop order \eqn{eq:eloss-prob-1} resums multiple scattering. It encompasses both the vacuum radiation that we have discussed in section~\ref{sec:c-coft-function} as well as the medium induced radiation component. Then, at one loop order we can write 
 \beq\label{eq:eloss-prob-1}
 S_1 (\epsilon,R)  &\approx &  \delta(\epsilon)+   \frac{g^2}{2 N_c}  \int \rmd^{d} k (2\pi) \delta(k^2)\theta(k^+) \delta(\epsilon - k^+ )\Theta_{\rm alg}(|\bk|-Rk^+/2) \nn
 && \times\left[k^+\frac{\rmd I}{\rmd k^+ \rmd^2\bk} \delta(\epsilon-k^+)-\delta(\epsilon) \int\rmd q^+ \frac{\rmd I}{\rmd q^+ \rmd^2\bk}\right] \,,
 \eeq
where the radiative spectrum reads
\beq 
k^+\frac{\rmd I}{\rmd k^+ \rmd^2\bk}&=&2 {\rm Re} \int _0^{+\infty}  \rmd {z^+_2}   \int_0^{z^+_2} \rmd z^+_1 \cP(+\infty, z_2^+; \bk,\bq)  \cK(z_2^+, z_1^+;\bq,\bk_1)\,.
\eeq

We can further extract the vacuum contribution that lives near the lower bound of the $z^+$ integration 
\beq
\frac{\rmd I}{\rmd k^+ \rmd^2\bk} = \frac{\rmd I_\vac}{\rmd k^+ \rmd^2\bk}+\frac{\rmd I_\med}{\rmd k^+ \rmd^2\bk}\,,
\eeq
where
\beq 
k^+\frac{\rmd I_\vac}{\rmd k^+ \rmd^2\bk} = \frac{\abar }{\bk^2}\,,
\eeq
and 
\begin{align}
k^+\frac{\rmd I_\med}{\rmd k^+ \rmd^2\bk}=2 {\rm Re} \int _0^{+\infty}  \rmd {z^+_2}   \int_0^{z^+_2} \rmd z^+_1 &\left[\cP(+\infty, z_2^+; \bk,\bq)  \cK(z_2^+, z_1^+;\bq,\bk_1)\right.  \nonumber\\
&\left. - \cP_0(+\infty, z_2^+; \bk,\bq) \cK_0(z_2^+, z_1^+;\bq,\bk_1)\right]\,.
\end{align}

Performing the medium average requires specifying a model for the medium correlations~\cite{Wang:1992qdg,Aurenche:2002pd}. The standard treatment employs the independent-scattering approximation, which assumes that the medium correlation length is much shorter than the mean free path. Within this kinetic description, the background field is modeled as a Gaussian random field, so that all higher-point correlators reduce to products of two-point functions and the medium average can be evaluated analytically. This approximation is briefly reviewed in appendix~\ref{app:field-correlator}.

Analytic expressions in both the single-hard-scattering and multiple-soft-scattering limits have been extensively discussed in the literature. These results form the basis of the GLV opacity expansion~\cite{Gyulassy:2000er,Wiedemann:2000za} and the BDMPS-Z description of medium-induced energy loss~\cite{Baier:1996kr,Baier:1996sk,Zakharov:1996fv,Zakharov:1997uu}. More recently, the Improved Opacity Expansion (IOE) was developed to interpolate systematically between these two regimes and provide analytic control beyond the multiple-soft-scattering approximation~\cite{Mehtar-Tani:2019tvy,Mehtar-Tani:2019ygg,Barata:2020sav,Barata:2021wuf}. Medium-induced radiation from a color antenna, including interference and color-decoherence effects, was originally studied in Refs.~\cite{Mehtar-Tani:2010ebp,
Mehtar-Tani:2011hma,
Mehtar-Tani:2011vlz,
Casalderrey-Solana:2011ule,
Mehtar-Tani:2011lic,
Mehtar-Tani:2012mfa} and has recently been extended beyond the harmonic-oscillator approximation within the IOE framework~\cite{Kuzmin:2025fyu}. We refer the reader to these works for detailed derivations and collect the expressions relevant in the single scattering approximation in appendix~\ref{app:glv}.

\section{Conclusion and outlook}

In this work, we have developed a factorization framework for inclusive jet production near threshold that provides a unified description of vacuum-like soft-collinear radiation, medium-induced energy loss, and color decoherence, building on the factorization approach \cite{Mehtar-Tani:2025xxd,Mehtar-Tani:2024smp,Vaidya:2026yfa}. In the limit $N\gg 1$, where corrections suppressed by powers of $1/N$ are neglected, the jet function refactorizes into collinear functions and soft-collinear operators constructed from light-like Wilson lines along the directions of the resolved partons inside the jet. The renormalization-group evolution of these operators resums logarithms of the ratio between the jet energy and the characteristic energy carried by out-of-cone radiation,
\beq
\alpha_s \ln\left(\frac{E}{E_{\rm loss}}\right)
\sim \alpha_s \ln N ,
\eeq
and reduces to the BMS equation in the large-$N_c$ limit. Under the parametric hierarchy between the time scales associated with vacuum-like radiation and medium-induced interactions, the ultraviolet evolution is unchanged by the medium. Medium-induced energy loss and color decoherence are instead encoded in the boundary condition of the soft-collinear evolution.

This formulation exposes a direct connection between color decoherence in a QCD medium and gluon saturation at small-$x$. The soft-collinear operators play a role analogous to dipole, quadrupole, and higher-point operators in high-energy factorization, while the medium coherence angle separates configurations that remain unresolved by the medium from those whose constituents lose energy as independent color charges. BMS evolution generates an energy dependence of this angular scale,
\beq
\theta_c(\tau)\sim \theta_c(0)\, e^{-c\tau},\qquad 
\qquad
\tau\sim\alpha_s\ln\left(\frac{E}{E_{\rm loss}}\right),
\eeq
which mirrors the asymptotic evolution of the saturation scale under BK evolution. For a dense and extended medium, the initial coherence angle is parametrically given by
\beq
\theta_c(0)\sim \left(\hat q L^3\right)^{-1/2}.
\eeq
More generally, the relevant initial angular scale is determined by the mechanism that resolves the soft radiation, with the jet radius $R$ providing the corresponding angular boundary in the vacuum.

The resulting evolution is nonlinear, reflecting the increasing number of resolved color sources produced during the shower. Using a simplified model related to the logistic equation, we have illustrated how these nonlinear corrections can moderate the growth of jet suppression at high-$p_T$. Although this model is intended as a qualitative illustration, it demonstrates how saturation-like dynamics may become phenomenologically relevant for jet quenching. A quantitative determination of this effect will require solving the full BMS evolution with a realistic medium-modified boundary condition. A concise account of the correspondence between gluon saturation and color decoherence established in this work, along with its phenomenological implications for jet quenching, is presented in the companion Letter~\cite{Caucal:2026}.

We have also shown how known results for medium-induced radiation from a single parton and from a color antenna emerge within a background-field effective theory formulated in the same language as vacuum QCD factorization. Multiple interactions with the medium are resummed into background-field Wilson lines and dressed propagators, while the expansion in soft-collinear quantum fluctuations generates the radiative corrections. This common theoretical formulation provides a natural starting point for a systematic assessment of perturbative uncertainties and for the construction of systematically improvable calculations.

Indeed, an important next step is the extension of the framework to higher perturbative accuracy. This requires computing the soft-collinear operators, the collinear functions, their renormalization-group evolution, and the medium-modified boundary conditions at the corresponding order. Since the vacuum BMS evolution is known up to three loops~\cite{Caron-Huot:2016tzz}, several ingredients can be imported directly from the existing literature. The remaining challenge is to determine the medium-dependent matching conditions with comparable precision.

Finally, the framework can be extended to less inclusive observables, including jet-substructure observables and energy--energy correlators. Such measurements introduce additional kinematic scales and may require further refactorization and resummation. Nevertheless, their dominant medium modification is often controlled by large-angle energy loss combined with the steeply falling jet spectrum, commonly referred to as the bias effect. Because this mechanism is naturally incorporated in the present formalism, our results provide a foundation for systematically improvable factorization theorems for a broad class of jet-quenching observables.

\section*{Acknowledgements}

P.~C. is funded by the Agence Nationale de la Recherche under
grant ANR-25-CE31-5230 (TMD-SAT). This work was supported by the U.S. Department of Energy under Contract No. DE-SC0012704. We are grateful for the support of the Saturated Glue (SURGE) Topical Theory Collaboration, funded by the U.S. Department of Energy, Office of Science, Office of Nuclear Physics. 
\appendix

\section{Useful integrals}
\label{app:dimreg-integrals}

For completeness, we collect below the integrals over the light-cone momentum $q^+$ and the rescaled angular variable ($\boldsymbol{\theta}\equiv 2\boldsymbol{q}/(q^+R)$) that enter the one-loop calculation of the soft-collinear function in Sec.~\ref{sec:one-loop-soft-function}. We have:
\begin{align}
\int_0^{+\infty} \frac{\rmd q^+}{(q^+)^{1+2\epsilon}}\left(\rme^{-q^+ \nu}-1\right) &=  \nu^{2\epsilon} \int_0^{+\infty} \rmd t \, t^{-1-2\epsilon}\, (\rme^{-t}-1) = \nu^{2\epsilon}\, \Gamma(-2\epsilon)\,,\\
&= -\nu^{2\epsilon}\left[\frac{1}{2\epsilon}+ \gamma_E + \left(\gamma_E^2+\frac{\pi^2}{6}\right) \,\epsilon+\cO(\epsilon^2)  \right]\,,
\end{align}
and 
\begin{align}
 \int^{+\infty}_{1} \frac{\rmd^{2-2\epsilon}  \btheta_\perp}{(2\pi)^{2-2\epsilon} \btheta^2} &=  \frac{1}{(2\pi)^d } \frac{ \pi^{d/2}}{\Gamma(d/2)}  \int_1^\infty \rmd \theta^2 (\theta^2)^{(d-4)/2}\,,\\
  &=  \frac{1}{(4\pi)^{1-\epsilon} \Gamma(1-\epsilon)}  \int_1^\infty \rmd \theta^2 (\theta^2)^{-\epsilon-1}\,,\\
   &=  \frac{1}{(4\pi)^{1-\epsilon} \Gamma(1-\epsilon)} \frac{1}{ \epsilon}\,,\\
    &= \frac{1}{4\pi} \left[\frac{1}{\epsilon}+\ln(4\pi)-\gamma_E+\left( \frac{1}{2}(\ln(4\pi)-\gamma_E)^2-\frac{\pi^2}{12}\right)\, \epsilon + \cO(\epsilon^2) \right]\,.
 \end{align}

\section{Factorization of multiple-interactions }\label{app:field-correlator}
The radiation rate is a fully non-perturbative quantity. However, one may factorize further owing to the fact that the in-medium correlation length is much shorter than the medium length.  More precisely, we have the following hierarchy, 
\beq
\xi_D \ll t_f \lesssim L  \,,
\eeq
where $\xi_D$ is the Debye screening length and $t_f$ the gluon coherence time. 

Integrating out the modes at distance scales of order the Debye length using HTL allows us to treat the Glauber gluon interactions as quasi-instantaneous and perturbatively over a distance $\xi_D$
\beq
 \frac{1}{d_R}{\rm tr}_c \langle A_{\bkg}^{\mu}(q) A_{\bkg}^{\ast,\nu}(q') \rangle =   (2\pi)^4\delta(q-q')  \rho^{\mu\nu}(q)\,,
\eeq
or 
\beq
 \frac{1}{d_R}{\rm tr}_c \langle A_{\bkg}^{-}(x^+, \bx) A_{\bkg}^{\ast,-}(y^+,\by) \rangle =  C_R\int \frac{\rmd^4 q} {(2\pi)^4} \rme^{-i \bq\cdot (\bx-\by)} \rme^{i q^-  (x-y)^+} \rho^{- - }(q) \,.
\eeq
Here the spectral density $\rho^{\mu\nu}(q)$ is to be computed using HTL. 
The Markovian approximation amounts effectively to replacing 
\beq
\rme^{i q^-  (x-y)^+}  \to (2\pi) \delta(x^+-y^+) \delta(q^- ) \,,
\eeq
yielding 
\begin{align}
 \frac{1}{d_R}{\rm tr}_c \langle A_{\bkg}^{-}(x^+, \bx) A_{\bkg}^{\ast,-}(y^+,\by) \rangle &\approx  C_R\delta(x^+-y^+)   \int \frac{\rmd^4 q} {(2\pi)^4} \rme^{-i \bq\cdot (\bx-\by)} \delta(q^-) \rho^{- - }(q) \,,\\
 &=  C_R \delta(x^+-y^+)   \int \frac{\rmd^2 \bq } {(2\pi)^2} \rme^{-i \bq\cdot (\bx-\by)}  C(\bq) \,,
\end{align}
where we have introduced the elastic rate
\beq
 C(\bq) =\int \frac{\rmd q^+ \rmd q^- } {(2\pi)^2}  (2\pi) \delta(q^-)  \rho^{- - }(q)  \, .
\eeq
This approximation is valid so long as the $k^-$ of the jet modes is much smaller than $q^-$ of the Glauber mode. This corresponds to the small $x$ approximation where both $k^-$ and $k^+$ momenta are strongly ordered. The case $k^-\sim q^-$ corresponds to $x\sim 1$ regime. 
\section{Dilute regime: Lipatov vertex }\label{app:glv}

Here we present the explicit expressions for the energy-loss distribution at NLO in the single-scattering approximation, corresponding to first order in the opacity expansion. This regime is commonly referred to as the GLV limit. We begin by recalling the general expression given in eq.\,\eqref{eq:eloss-prob-1}.
 \beq\label{eq:eloss-prob-1bis}
S^{(1)}_1 (\epsilon,R)  &\approx &   \frac{g^2}{2 N_c}  \int \rmd^{d} k (2\pi) \delta(k^2)\theta(k^+) \delta(\epsilon - k^+ )\Theta_{\rm alg} \nn
 && 2 {\rm Re}\int _0^{+\infty}  \rmd {z^+_2}   \int_0^{z^+_2} \rmd z^+_1 \cP(+\infty, z_2^+; \bk,\bq)  \cK(z_2^+, z_1^+;\bq,\bk_1)  \,.
 \eeq
 We proceed further to expand to leading order order in opacity. The zeroth order corresponds to the vacuum contribution
   \beq\label{eq:eloss-vac}
S^{(1,0)}_1 (\epsilon,R)  &\approx &   \frac{g^2}{2 N_c}  \int \rmd^{d} k (2\pi) \delta(k^2)\theta(k^+) \delta(\epsilon - k^+ )\Theta_{\rm alg} \nn
 &&2 {\rm Re} \int _0^{+\infty}  \rmd {z^+_2}   \int_0^{z^+_2} \rmd z^+_1 \cP_0(\bk-\bq)\cK_0(z_2^+, z_1^+;\bq,\bk_1) \,.
 \eeq
using the fact that 
\beq
\cP^{(0)}(\bk-\bq) =(2\pi)^2\delta(\bk-\bq) \,,
\eeq
and 
\beq
\cK^{(0)}(z_2^+, z_1^+;\bq,\bk_1)= \frac{N_c^2-1}{(k^+)^2} (2\pi)^2\delta(\bk_1-\bq)(\bq\cdot \bk_1)  \rme^{-i\frac{\bq^2}{2k^+} (z_2-z_1)^+}\,,
\eeq
a straightforward calculation yields
 \beq\label{eq:eloss-vac}
S^{(1,0)}_1 (\epsilon,R)  &\approx &   4 g^2 C_F  \int \rmd^{d} k (2\pi) \delta(k^2)\theta(k^+) \delta(\epsilon - k^+ )\Theta_{\rm alg} \frac{1}{\bk^2}\,.
 \eeq
 
 Let us turn to a single gluon radiation contribution at leading order in opacity. To do so, we need to expand \eqn{eq:eloss-prob-1bis} to leading order in opacity. We obtain two terms 
 \begin{align}\label{eq:eloss-prob-opacity1}
&S^{(1,1)}_1 (\epsilon,R)  \approx    \frac{g^2}{2 N_c}  \int \rmd^{d} k (2\pi) \delta(k^2)\theta(k^+) \delta(\epsilon - k^+ )\Theta_{\rm alg} \times2 {\rm Re}\int _0^{+\infty}  \rmd {z^+_2}   \int_0^{z^+_2} \rmd z^+_1  \nn
&  \times\left[   \cP^{(1)}(+\infty, z_2^+; \bq,\bk_1)  \cK^{(0)}(z_2^+, z_1^+;\bk,\bq)+\cP^{(0)}(+\infty, z_2^+; \bq,\bk_1)  \cK^{(1)}(z_2^+, z_1^+;\bk,\bq)   \right]\,,
 \end{align}
where 
\begin{align}
 \cK^{(0)}(z_2^+, z_1^+;\bq,\bk_1) &=(N_c^2-1) \frac{\bq^2}{(k^+)^2} (2\pi)^2\delta(\bq-\bk_1) \, \rme^{-i\frac{\bq^2}{2k^+}(z_2-z_1)^+}\,,\\ 
\cK^{(1)}(z_2^+, z_1^+;\bq,\bk_1) &=(N_c^2-1) C_A \frac{\bq\cdot\bk_1 }{(k^+)^2}   \int_{z^+_1}^{z^+_2} \rmd z_3^+ \rme^{-i\frac{\bq^2}{2k^+}(z_2-z_3)^+} C_+(\bq-\bk_1,z^+_3)  \, \rme^{-i\frac{\bk_1^2}{2k^+}(z_3-z_1)^+} \,,\\
 \cP^{(0)}(+\infty, z_2^+; \bk,\bq)&=  (2\pi)^2\delta(\bk-\bq) \,,\\
 \cP^{(1)}(+\infty, z_2^+; \bk,\bq) &=C_A  \int_{z^+_2}^{+\infty} \rmd z_3^+ C_+(\bq-\bk_1,z^+_3) \,,
\end{align}
and $C_+$ is defined by its action on a test function $f$ 
\beq
 \int \rmd \bq \, C_+(\bq) f(\bq) =  \int \rmd \bq\,  C(\bq) \, (f(\bq) -f(\0))\,.
\eeq
We can now carry out the integration over the light-cone times $z_1^+$ and $z_2^+$. We find
 \beq\label{eq:eloss-prob-opacity1-2}
&& S^{(1,1)}_1 (\epsilon,R)  \approx    g^2\, C_F  \int \rmd^{d} k (2\pi) \delta(k^2)\theta(k^+) \delta(\epsilon - k^+ )\Theta_{\rm alg}  \int_{z^+_2}^{+\infty} \rmd z_3^+ \nn
 && 8 C_A \left[   - \frac{1}{\bq^2}+ \frac{\bq\cdot \bk }{\bk^2 \bq^2 } \right] \left(1-\rme^{-i\frac{\bq^2}{2k^+} z_3^+}\right)\, C_+(\bk-\bq,z_3^+)\,,
 \eeq
or equivalently 
 \beq\label{eq:eloss-prob-opacity1-2}
&& S^{(1,1)}_1 (\epsilon,R)  \approx     g^2\, C_F  \int \rmd^{d} k (2\pi) \delta(k^2)\theta(k^+) \delta(\epsilon - k^+ )\Theta_{\rm alg}  \int_{0}^{+\infty} \rmd z_3^+ \nn
 && 8 C_A  \,  \frac{\bq\cdot \bk }{\bk^2 (\bk-\bq)^2 } \,  \left[1-\cos\left(\frac{(\bk-\bq)^2}{2k^+} z_3^+\right)\right]\, C(\bq,z_3^+)\,.
 \eeq
 
 It is instructive to explore the large medium length limit or more precisely, the limit of short formation time, i.e., $2k^+/(\bk-\bq)^2 \ll L$. We have
  \begin{align}
  8 C_A  \,  \frac{\bq\cdot \bk }{\bk^2 (\bk-\bq)^2 }&= 4 C_A  \, \ \frac{\bq^2 + \bk^2 -(\bk-\bq)^2}{\bk^2 (\bk-\bq)^2 } \,,\\
  & = 4 C_A  \, \ \left[\frac{\bq^2 }{\bk^2 (\bk-\bq)^2 }  +\frac{1}{(\bk-\bq)^2} -\frac{1}{\bk^2}\right] \,.
 \end{align}
 Hence, the real gluon emission contribution to the energy loss probability distribution at leading order in opacity reads
  \begin{align}\label{eq:eloss-prob-opacity1-L}
 S^{(1,1)}_1 (\epsilon,R) & \approx     g^2\, C_F  \int \rmd^{d} k (2\pi) \delta(k^2)\theta(k^+) \delta(\epsilon - k^+ )\Theta_{\rm alg}  \int_{0}^{+\infty} \rmd z_3^+ \nonumber\\
 &\times 4 C_A  \,  \ \left[\frac{\bq^2 }{\bk^2 (\bk-\bq)^2 }  +\frac{1}{(\bk-\bq)^2} -\frac{1}{\bk^2}\right] \, C(\bq,z_3^+)\,.
 \end{align}
The first term stands for the Lipatov vertex squared while the second and third terms correspond to the broadening of a vacuum radiation.

\bibliographystyle{JHEP}
\bibliography{references.bib}

\end{document}